\documentclass[usenatbib]{mn2e}

\pdfoutput=1

\usepackage{longtable}
\usepackage{url}
\usepackage{fixltx2e}
\usepackage{mathptmx}
\usepackage[pdftex]{graphicx}
\usepackage{color}

\newcommand{\kpc}{\rm\thinspace kpc}

\newcommand{\km}{\rm\thinspace km}

\newcommand{\cm}{\rm\thinspace cm}
\newcommand{\yr}{\rm\thinspace yr}

\newcommand{\s}{\rm\thinspace s}

\newcommand{\Msun}{\hbox{$\rm\thinspace M_{\odot}$}}

\newcommand{\Msunpyr}{\hbox{$\Msun\yr^{-1}\,$}}

\newcommand{\keV}{\rm\thinspace keV}

\newcommand{\kmps}{\hbox{$\km\s^{-1}\,$}}

\newcommand{\Zsun}{\hbox{$\thinspace \mathrm{Z}_{\odot}$}}

\newcommand{\psqcm}{\hbox{$\cm^{-2}\,$}}

\newcommand{\ps}{\hbox{$\s^{-1}\,$}}

\begin{document}

\title[Turbulence in galaxy clusters] {Constraints on turbulent
  velocity broadening for a sample of clusters, groups
  and elliptical galaxies using \emph{XMM-Newton}}

\author
[J.S. Sanders et al]
{J.~S. Sanders$^1$,
  A.~C. Fabian$^1$ and R.~K. Smith$^2$
  \\
  $^1$ Institute of Astronomy, Madingley Road, Cambridge. CB3 0HA\\
$^2$ MS 6, 60 Garden Street, Harvard-Smithsonian Center
for Astrophysics, Cambridge, MA 02138, USA
}
\maketitle

\begin{abstract}
  Using the width of emission lines in \emph{XMM-Newton} Reflection
  Grating Spectrometer spectra, we place direct constraints on the
  turbulent velocities of the X-ray emitting medium in the cores of 62
  galaxy clusters, groups and elliptical galaxies.  We find five
  objects where we can place an upper limit on the line-of-sight
  broadening of $500\kmps$ (90 per cent confidence level), using a
  single thermal component model. Two other objects are lower than
  this limit when two thermal components are used.  Half of the
  objects examined have an upper limit on the velocity broadening of
  less than $700\kmps$.  To look for objects which have significant
  turbulent broadening, we use \emph{Chandra} spectral maps to compute
  the expected broadening caused by the spatial extent of the
  source. Comparing these with our observed results, we find that
  Klemola\,44 has extra broadening at the level of
  $1500\kmps$. RX\,J1347.5-1145 shows weak evidence for turbulent
  velocities at $800\kmps$.  In addition we obtain limits on
  turbulence for Zw\,3146, Abell~496, Abell~1795, Abell~2204 and
  HCG~62 of less than $200\kmps$. After subtraction of the spatial
  contribution and including a $50\kmps$ systematic uncertainty, we
  find at least 15 sources with less than 20 per cent of the thermal
  energy density in turbulence.
\end{abstract}


\section{Introduction}
Measurements of the velocity structure of the intracluster medium
(ICM) are of great interest. They are important for measuring the
turbulence predicted to be injected into clusters from mergers or the
accretion of material, tracing how the central nucleus injects energy
into its surroundings, examining ICM transport properties, such as
looking at metal diffusion or sound waves, and for the determination
of cluster gravitational potentials using X-ray observations.
However, there are few observational constraints on the amount of bulk
flow or random motions within the intracluster medium.

Theoretical models of the intracluster medium have predicted that the
fraction of pressure support in gas motions is typically 5 to 15 per
cent (e.g. \citealt{Lau09,Vazza09}). Three-dimensional hydrodynamic
simulations typically include at most numerical viscosity, and most do
not attempt to examine the effects of magnetic fields on gas
motions. Some simulations do however attempt to model the effect of
the central active nucleus on the gas motions. \cite{Bruggen05}
predict motions in the range $500-1000\kmps$ around the central
nucleus. \cite{Heinz10} finds cluster-wide turbulent line-of-sight
velocities of $\sim 500 \kmps$ in their simulations of clusters
containing AGN jets, examining how easily these signals would be
detected by \emph{IXO}.

Turbulent motions have been used to explain the lack of cool X-ray
emitting gas in the cores of galaxy clusters. Random motions of
$50-150\kmps$ could randomize the magnetic field direction, enhancing
conduction and helping suppress central cooling
\citep{Ruszkowski10}. Turbulence should trigger plasma instabilities,
giving thermally stable local heating rates comparable to the cooling
rates in the intracluster medium \citep{Rosin10,Kunz10}.

Observational constraints on random turbulent motions in galaxy
clusters are mostly indirect in nature. In the elliptical galaxy
NGC\,4636, \cite{Xu02} examined the strength of the resonantly
scattered 15{\AA} Fe~\textsc{xvii} line relative to the line at
{17.1\AA}. The ICM can be optically thick in the light of resonant
lines given correct temperatures and sufficient densities. Random
motions decrease the effect of resonant scattering because the
resonant lines are velocity-broadened. \cite{Xu02} observed the
consequences of resonant scattering, inferring that the turbulent
velocity dispersion is less than 10 per cent of the sound
speed. \cite{Werner09} also examined this object, concluding a maximum
of 5 per cent of energy is in turbulent motions.

Random motions contribute to nonthermal ICM pressure.
\cite{Churazov08} derived the the gravitational mass profiles from
optical and X-ray data in M87 and NGC\,1399. Stars act as collisionless
particles, so comparing their potentials from that derived from X-ray
data can measure or place limits on the nonthermal pressure.  They
obtained an upper limit on the nonthermal pressure of $\sim 10$ per
cent of the thermal gas pressure.

One of the few constraints on turbulence in an unrelaxed galaxy
cluster was made by \cite{Schuecker04}, who examined the power
spectrum of pressure fluctuations in the Coma cluster. They deduced
that a minimum of 10 per cent of the total pressure in Coma is in the
form of turbulence.

The shape of X-ray emission lines provides information about velocity
structure in clusters \citep{Inogamov03}. The low resolution spectra
from CCD instruments can be used to look for changes in bulk velocity
as a function of position, if the gain of the instrument is stable
enough. Such flows were reported using \emph{ASCA} and \emph{Chandra}
in Centaurus \citep{DupkeBregman01,DupkeBregman06} and other clusters
\citep{DupkeBregman05}. However, \cite{Ota07} placed an upper limit on
bulk flows of $1400\kmps$ in the Centaurus cluster using
\emph{Suzaku}.

The only direct limit on turbulence in the X-ray waveband came from
our recent work examining the \emph{XMM-Newton} Reflection Grating
Spectrometer (RGS) spectra from a long observation of the X-ray bright
galaxy cluster Abell 1835 \citep{Sanders10_A1835}. Since this cluster
is at $z=0.2523$ and has a compact cool core, the line emission is
concentrated in a small region on the sky. The broadening of the
emission lines by the slitless RGS spectrometers is therefore
small. We were able to place an upper limit on the total broadening,
including the turbulent component, of $274 \kmps$, at the 90 per cent
level. The ratio of turbulent to thermal energy density in the core of
Abell\,1835 is less than 13 per cent.

In this paper, we continue this work by examining RGS spectra of the
most point-like clusters, groups and elliptical galaxies in the
\emph{XMM-Newton} archive, to obtain the best current direct limits on
turbulent velocity broadening of emission lines.

All uncertainties are at the $1\sigma$ level, unless stated
otherwise. We use the Solar relative abundances of
\cite{AndersGrevesse89}.

\section{Analysis}
In the first part of our analysis we measure conservative upper limits
on the turbulence in our sample by assuming that the objects are point
sources. We measure the total line width of the sources in \kmps,
which includes the turbulent component.

As the RGS instruments are slitless spectrometers and the examined
objects are not point sources, the spectra are broadened because of
the spatial extent of the source in the dispersion direction.  The
effect of the broadening of the spectrum by the spatial extent of the
source is given by
\begin{equation}
  \Delta \lambda \approx \frac{0.124}{m} \Delta \theta \: \textrm{\AA},
\end{equation}
where $m$ is the spectral order and $\Delta\theta$ is the half energy
width of the source in arcmin \citep{Brinkman98}.  This broadening is
because when we measure a particular dispersion angle on the detector
for an X-ray photon, we cannot differentiate between a change in
wavelength and a change in position along the dispersion
direction. However, there is a low resolution measurement of the
energy of each event by the CCD detector which is used to
differentiate the multiple spectral orders, also removing noise and
signal from the calibration sources.

The observed broadening of emission lines is the sum of this spatial
broadening, the thermal broadening of the ICM and any turbulent
motions within the observed region.  If we determine what the upper
bound is to the broadening of the emission lines, beyond the
instrumental point source broadening, subtracting the thermal
broadening, we can place an upper limit on the turbulent broadening of
the source.  In addition we can obtain measurements of the redshift of
the objects by measuring the emission line centroids.

It would be useful to be able to remove the broadening due to the
source extent. To do this we need to know how the high resolution
spectrum of the source varies along the dispersion direction, which is
not known. We can estimate how it varies by examining maps of the
source properties (temperature, metallicity and abundance) constructed
using spatially-resolved spectroscopy of \emph{Chandra} data. These
data have much lower spectral resolution than the RGS data, so cannot
remove the degeneracies completely. As the RGS gratings and
\emph{Chandra} detectors are fundamentally different kinds of
instrument, with potential calibration uncertainties, a joint analysis
is difficult.

We therefore construct synthetic RGS spectra from \emph{Chandra}
spectral maps of several of the sources in the sample. These spectra
do not contain any turbulent emission line broadening, but do include
spatial and thermal broadening. By measuring the width of the emission
lines in these theoretical spectra, and comparing them with the width
of the lines in the real spectra, we can identify objects in which
there is likely to be significant turbulent velocities.

\subsection{Sample selection}
The objects we examined were selected on the basis of strong emission
lines in order to get good limits on broadening. We examined by eye
all the clusters in the \emph{XMM-Newton} archive which came under the
category `groups of galaxies, clusters of galaxies, and superclusters'
and ellipticals in `galaxies and galactic surveys'. We looked for
those clusters, ellipticals or groups which had a bright central peak
in the EPIC images and emission lines in the preview RGS spectra or
when displayed with the \emph{XMM-Newton} online BiRD (Browsing
Interface for RGS Data) interface, or gas temperatures where there
should be emission lines. The sample is therefore not statistically
complete or rigorous and is biased towards bright relaxed objects.

We list the objects in Table \ref{tab:obj} and their redshifts taken
from the NED database. As the absolute energy calibration for the RGS
instruments depends on the source position being correct, we use the
X-ray position measurements using \emph{Chandra} if possible. We use
the centroid position from the ACCEPT cluster archive
\citep{Cavagnolo09} if available, verifying the position from
\emph{Chandra} data manually to find the X-ray peak. For those
clusters without \emph{Chandra} observations we used the \emph{XMM}
EPIC peak position.

\subsection{Spectral extraction}
We downloaded the datasets for each of the clusters listed in
Table~\ref{tab:obj} from the \emph{XMM} archive. We ignored some
datasets which were very short or where the lightcurve showed strong
flaring. The datasets were processed individually through the RGS
\textsc{rgsproc} pipeline (version 1.28.3, part of \textsc{sas}
10.0.0). We excluded time periods where the count rate for events with
flag values of 8 or 16, on CCD 9, and an absolute cross-dispersion
angle of $1.5\times 10^{-4}$ (the parameters used in
\citealt{Tamura03Calib}), was greater than $0.2\ps$.

\begin{figure}
  \includegraphics[width=\columnwidth]{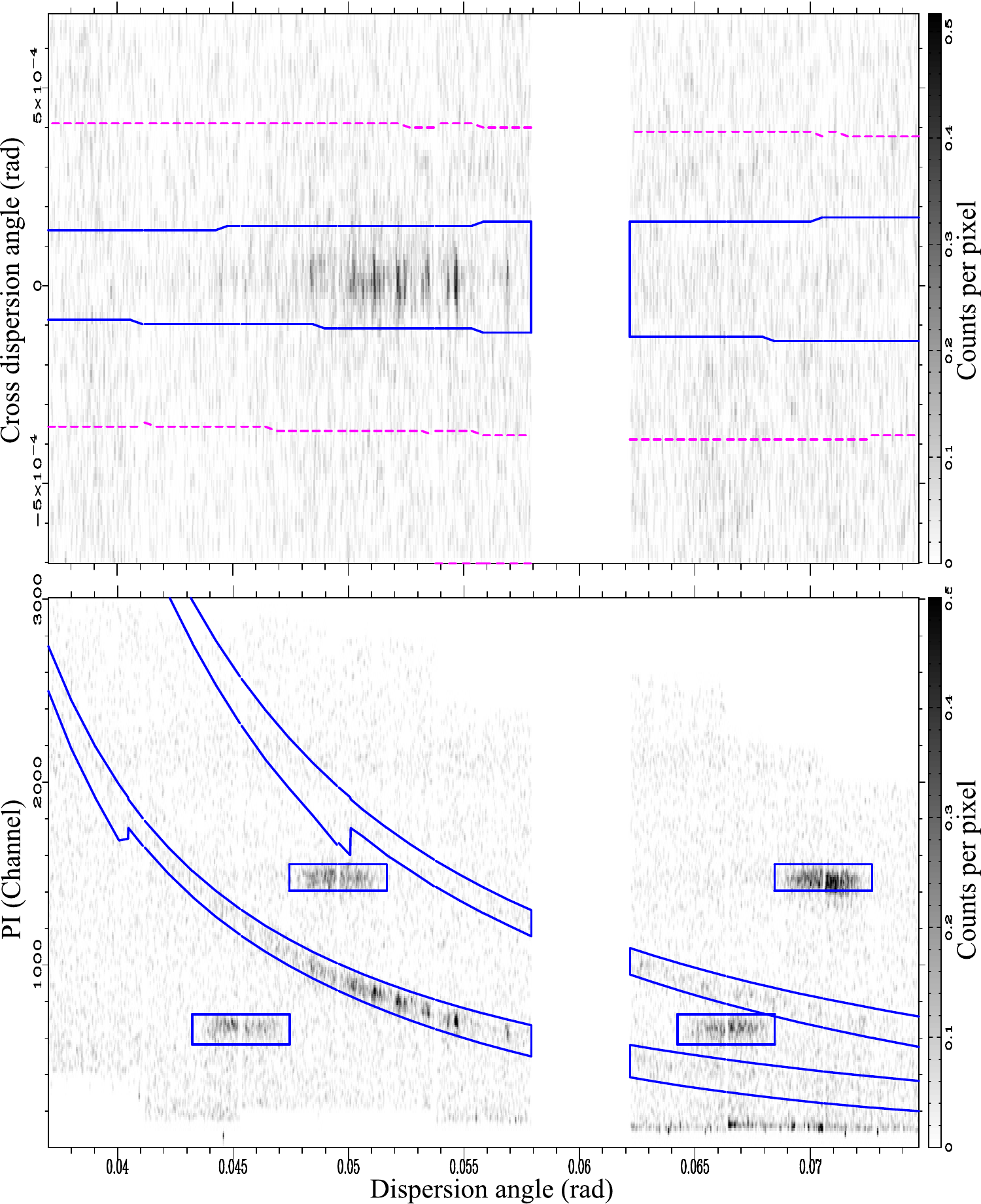}
  \caption{Views of the NGC\,4261 0502120101 RGS2 dataset. The bottom
    panel shows the X-ray events, plotting the CCD-measured energy
    (PI) against the the dispersion angle on the detector. Shown are
    the first and second order selection regions (corresponding to 90
    per cent of the pulse height distribution) and the regions for the
    calibration sources. The top panel shows those events in the first
    order spectrum, as a function of dispersion angle and
    cross-dispersion angle. Within the solid lines are the source
    extraction region (corresponding to 90 per cent of the width of a
    point source) and outside the dashed lines are the background
    extraction region.}
  \label{fig:banana}
\end{figure}

Foreground spectra were extracted from within 90 per cent of the PSF
(point spread function) of a point source, including 90 per cent of
the peak of the pulse-height distribution (this is the CCD measurement
of energy). The selection region roughly corresponds to a 50~arcsec
wide strip across the centre of the cluster. We created spectra for
background subtraction from the region outside of 98 per cent of the
PSF. Shown in Table~\ref{tab:obj} are the total cleaned exposure times
(summing the RGS1 and RGS2 exposure times) and average foreground and
background count rates.

The wavelength binning option was used to enable spectral binning by
wavelength so that the spectra from the two RGS instruments could be
combined. For a particular spectral order and object, we combined the
the RGS1 and RGS2 spectra, responses and background files from the
appropriate datasets.

We show in Fig.~\ref{fig:banana} the spectral extraction regions for a
line-rich bright source, NGC 4261. The bottom panel shows the
dispersion angle of each X-ray event on the RGS2 detector plotted
against the PI (pulse invariant) value, which is a measure of the
energy of the X-ray photon. The two spectral orders are split up by
the pulse height distribution selection. Also seen are the calibration
source events, excluded using the rectangles shown. The top plot shows
the distribution of the first order events as a function of
cross-dispersion direction and dispersion angle.

\begin{figure}
  \includegraphics[width=\columnwidth]{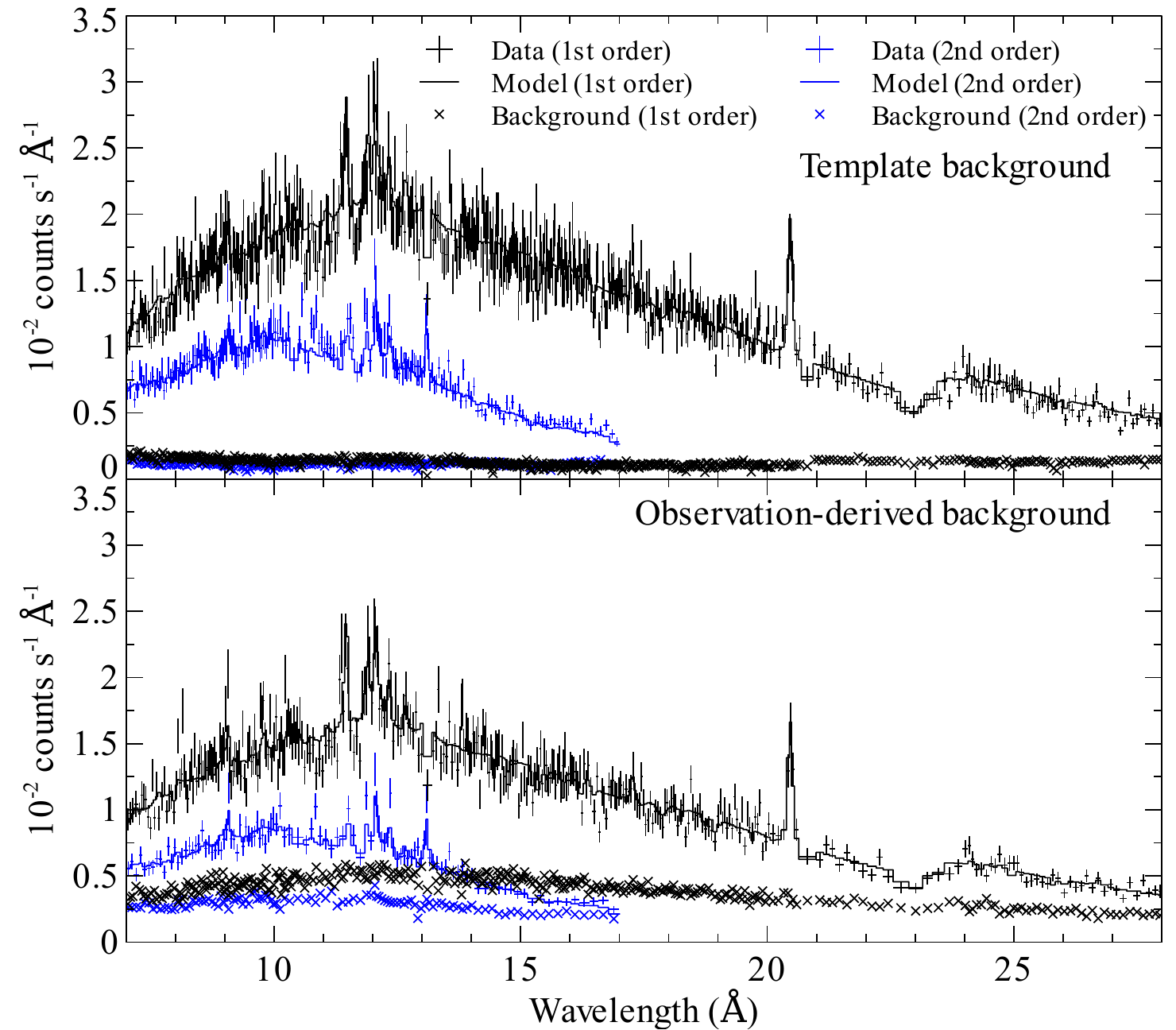}
  \caption{Comparison of background-subtracted data using backgrounds
    generated from templates (top panel) and from backgrounds from the
    observation itself (bottom panel) for the extended cluster Abell
    2029.}
  \label{fig:a2029bg}
\end{figure}

Some of the most extended objects have background spectra which are
contaminated by cluster emission. Fig.~\ref{fig:a2029bg} shows the
background subtraction for Abell~2029 with template backgrounds
generated using the \textsc{sas} \textsc{rgsbkgmodel} tool and with
backgrounds derived from the observation itself. For these extended
objects the spectrum from the background region is smoother than the
line emission from the centre. This is because this background
emission comes from a large part of the cluster, much larger than the
core, and is highly broadened.

For Abell~2029 the line width is identical when using a template
background or a background derived from the observation. For some
other objects, for example Centaurus, the line width is narrower by
$\sim 100\kmps$ using an observation-derived background because the
broader component of the emission lines has been subtracted.

\begin{figure}
  \includegraphics[width=\columnwidth]{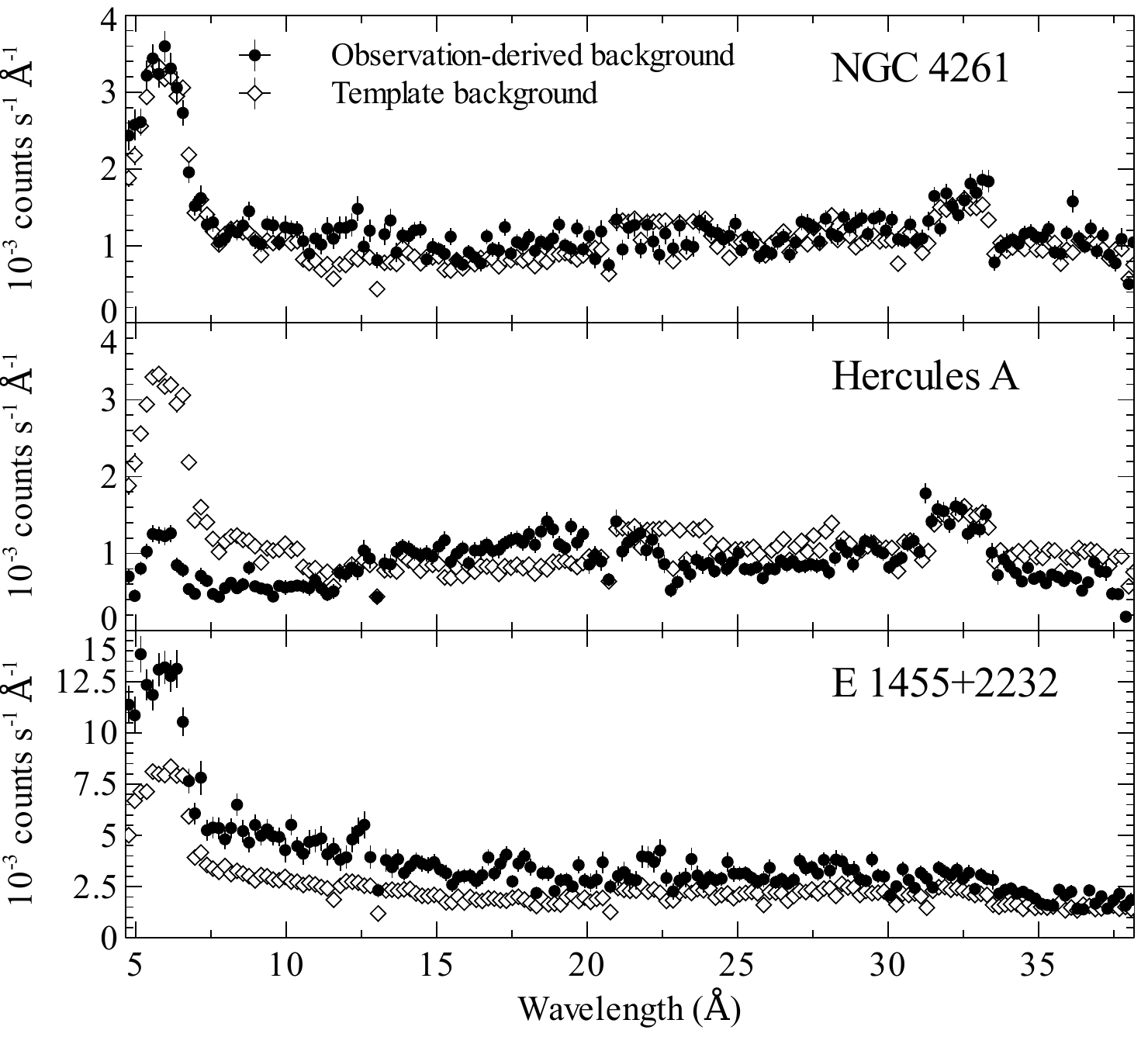}
  \caption{Comparisons of template and observation-derived backgrounds
    for three observations.}
  \label{fig:back_compar}
\end{figure}

We note that our background spectra for several compact objects do not
match the template backgrounds generated by the \textsc{sas}
\textsc{rgsbkgmodel} program, even at wavelengths where the RGS
instruments have little effective area.  Although some objects such as
NGC~4261 (Fig.~\ref{fig:back_compar}) match the template backgrounds
well, others such as Hercules A and E\,1455+2232 show substantial
mismatches between the templates and observed background at short
wavelengths.

\subsection{Spectral fitting}
\label{sect:specfitting}
For each object we simultaneously fit the spectra with an
\textsc{apec} 1.3.1 spectral model \citep{SmithApec01}. In the
spectral model we fit the temperature, Galactic absorbing column
density, emission measure and abundances for O, Ne, Mg, Si, Fe and
Ni. All other elemental abundances were fixed to the have the same
ratio relative to Solar as Fe, as their emission lines were weak in
the wavelength range examined. The lines in the spectral model were
given Gaussian line widths. The width of the lines were the
appropriate thermal line widths for an ion at the model temperature,
plus an additional velocity width added in quadrature, which was a
free parameter in the fits. This additional velocity parameter is of
most interest here, as it includes the turbulent broadening and
spatial broadening of the object. The redshift of the thermal model is
also a free parameter in the spectral fitting.

We simultaneously fit the first order spectra between 7 and {28\AA}
and second order spectra between 7 and {17\AA}, minimising the
\textsc{xspec} modified Cash statistic when fitting. We used
\textsc{xspec} version 12.6.0. Using C statistics rather than $\chi^2$
allows us to not bin or group the spectra, helping preserve the energy
resolution.

The spectral fitting results are shown in two tables. Table
\ref{tab:param} lists the best fitting temperature, column density and
metallicities. Table \ref{tab:vel} shows the best fitting redshift and
90 per cent upper limit on the line broadening, in $\kmps$.

Some of the cooler objects are extremely line dominated. For these
objects the continuum and emission measure are hard to determine. This
leads to large uncertainties on the metallicities, which are measured
relative to the continuum. Where the metallicity uncertainties are
large, we fix the Fe metallicity to Solar. The other metallicites and
emission measures become much better determined when this is done.
These objects are listed as having Fe equal to 1 in Table
\ref{tab:param}.

\begin{figure*}
  \includegraphics[width=\textwidth]{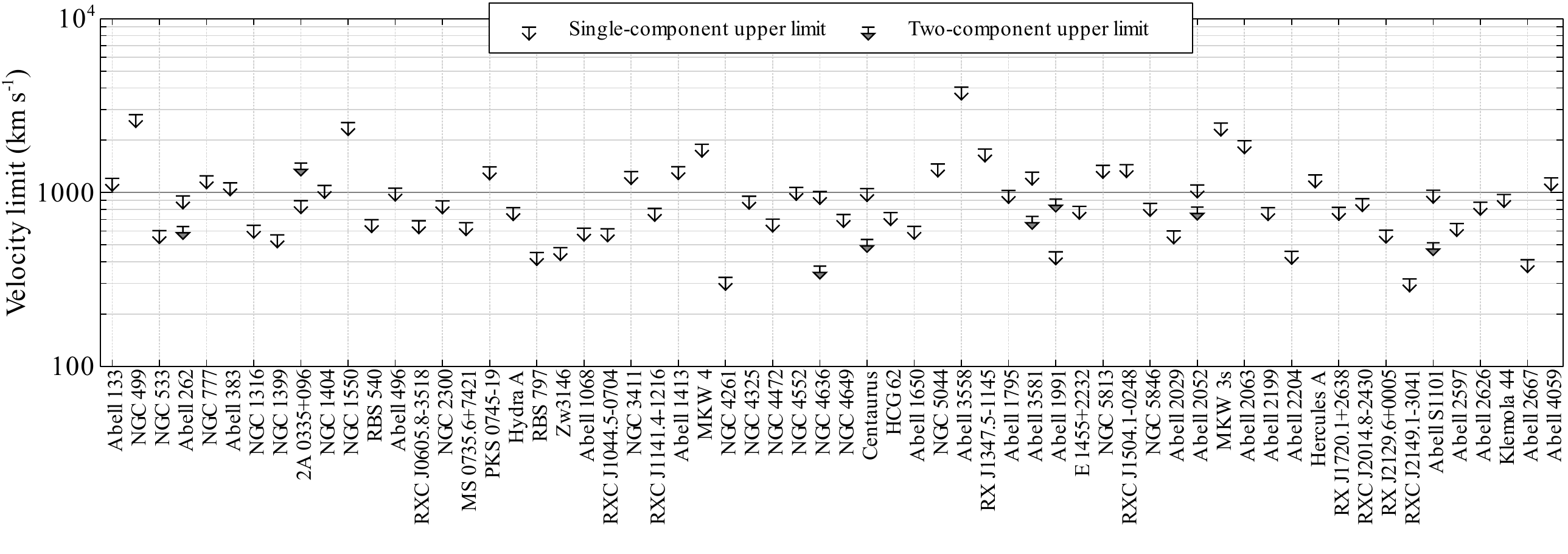}
  \caption{Upper limits (90 per cent) on the turbulent velocity
    broadening of the spectra. Shown are the limits for single thermal
    component modelling for all objects, and two temperature modelling
    for selected objects.}
  \label{fig:velocities}
\end{figure*}

We show the 90 per cent upper bound on the broadening of the emission
line in Fig.~\ref{fig:velocities}. These were calculated with the
\textsc{xspec} error command, which examines how the fit quality
changes when a model parameter is stepped over a range of values.

\begin{figure*}
  \includegraphics[width=\textwidth]{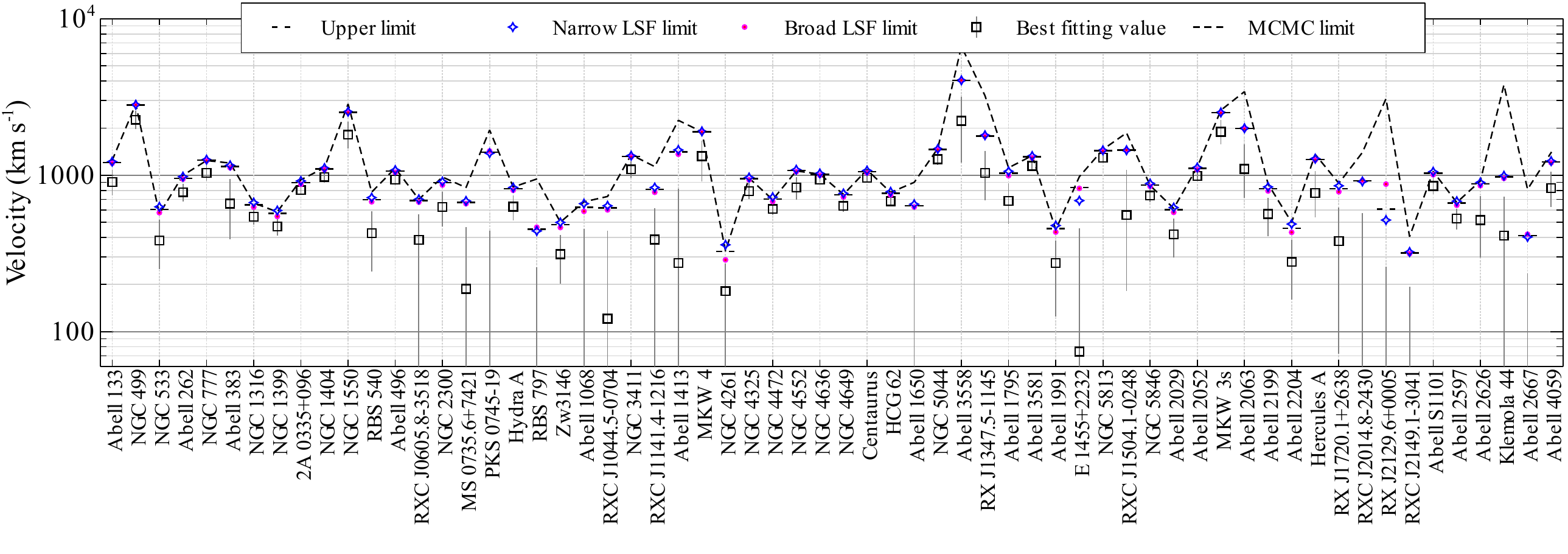}
  \caption{A comparison between the upper limits on the line widths
    obtained with the standard calibration, and responses in which the
    LSF is 10 per cent narrower or 10 per cent broader than
    standard. The continuous line shows the upper limits derived from
    the MCMC analysis. Also plotted are the best fitting line width
    and $1\sigma$ error bars, which include the spatial component of
    broadening.}
  \label{fig:velocities_xtra}
\end{figure*}

\begin{figure}
  \includegraphics[width=\columnwidth]{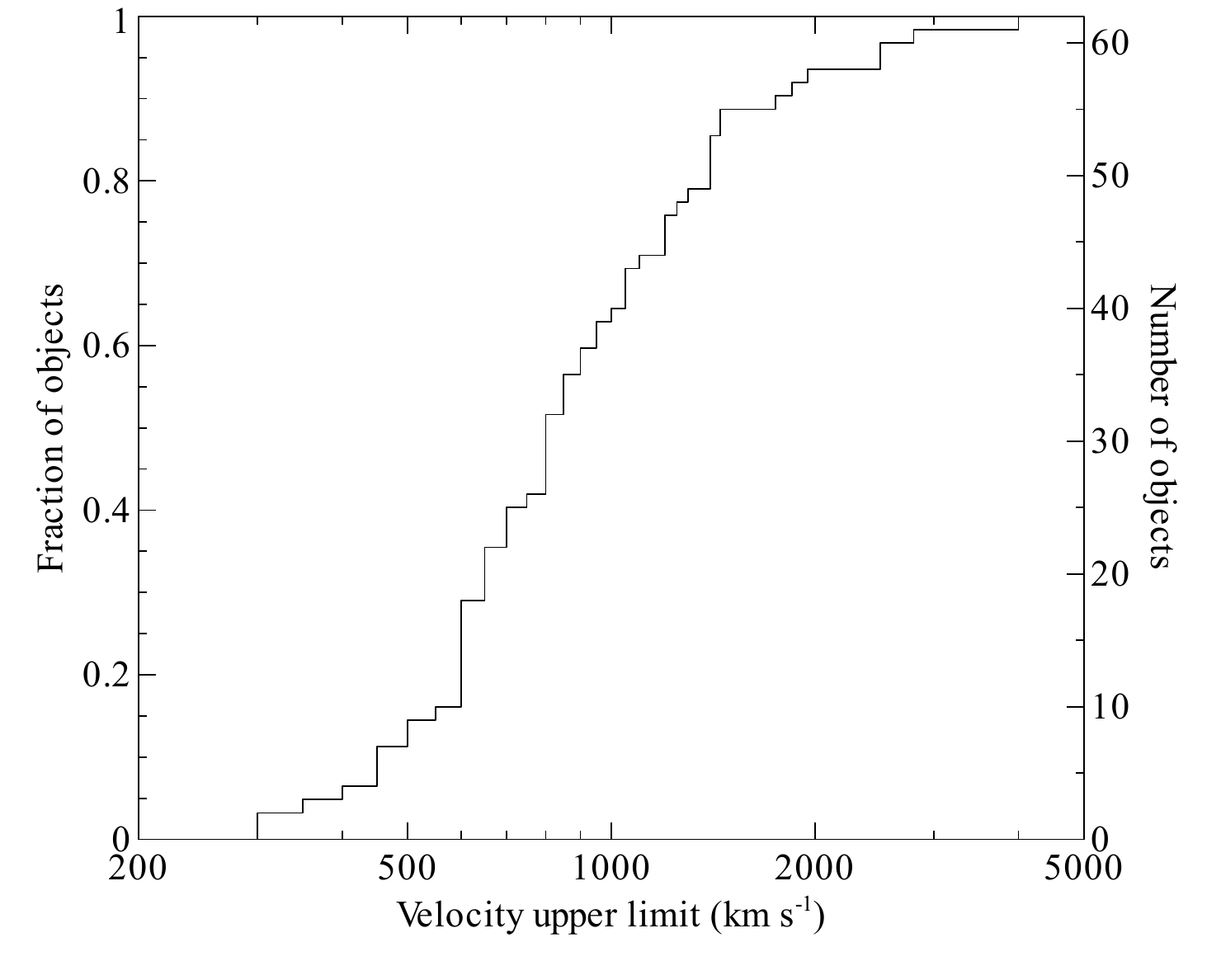}
  \caption{Cumulative distribution of upper limits on velocity
    broadening, using two temperature component results where
    appropriate. Note that our sample is highly biased, so this plot
    should not be used as any indication of the real velocity
    distribution within clusters or groups. The plot is to indicate
    the distribution of our results.}
  \label{fig:velcuml}
\end{figure}

\begin{figure}
  \includegraphics[width=\columnwidth]{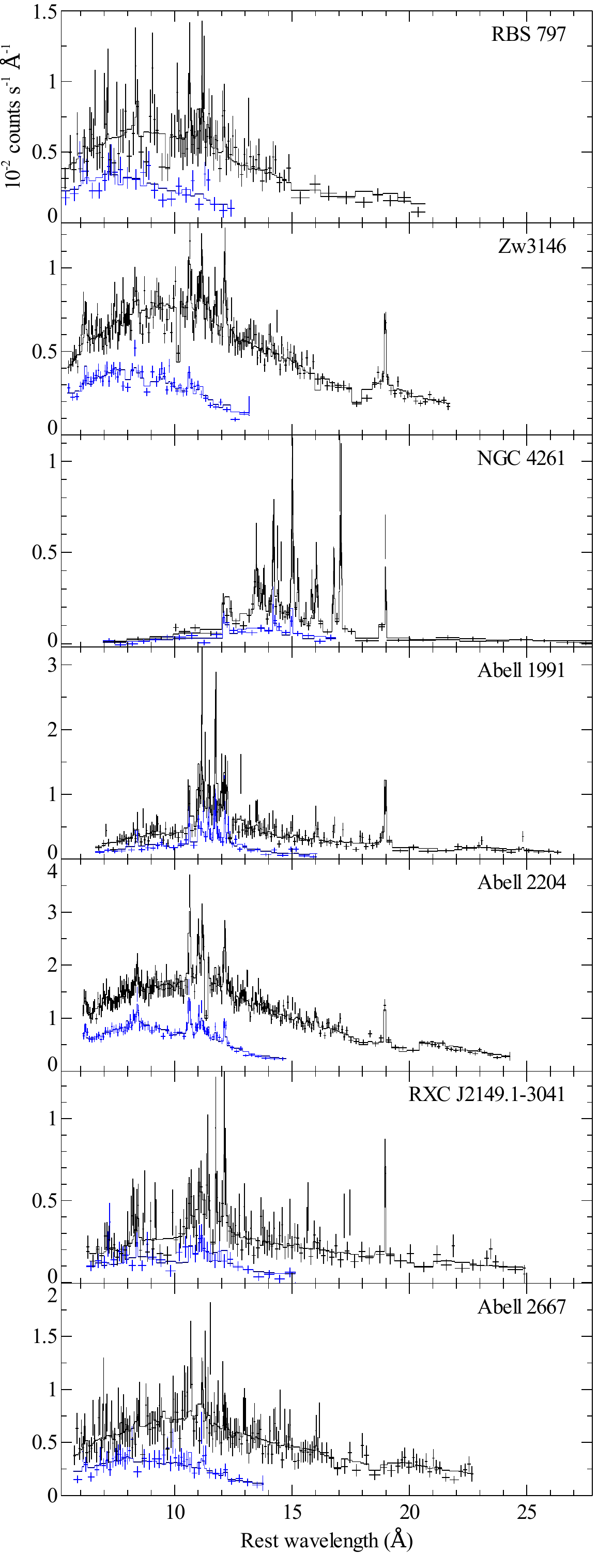}
  \caption{Rebinned combined RGS spectra and best fitting models for
    the objects where the upper limit on broadening of the emission
    lines is less than $500\kmps$. The upper and lower data points and
    models are for first and second order spectra, respectively.}
  \label{fig:spectralt500}
\end{figure}

These values are only upper limits because we do not know the effect
of the extent of the source in the measurements, which may be the
dominant contribution. To show whether the the upper limit is
primarily due to the quality of the spectrum or whether the width of
the emission line well determined, we also plot the best fitting
broadening and its $1\sigma$ error bars in
Fig.~\ref{fig:velocities_xtra}.  The cumulative distribution of
broadening upper limits is shown in Fig.~\ref{fig:velcuml}. Seven of
the objects have upper limits less than $500\kmps$. We show rebinned
spectra of these objects in Fig.~\ref{fig:spectralt500} with their
best fitting spectral models.  There are also two objects which give
limits better than $500\kmps$ for a second temperature component added
to the model (NGC~4636 and Abell~S1101).

\begin{figure}
  \includegraphics[width=\columnwidth]{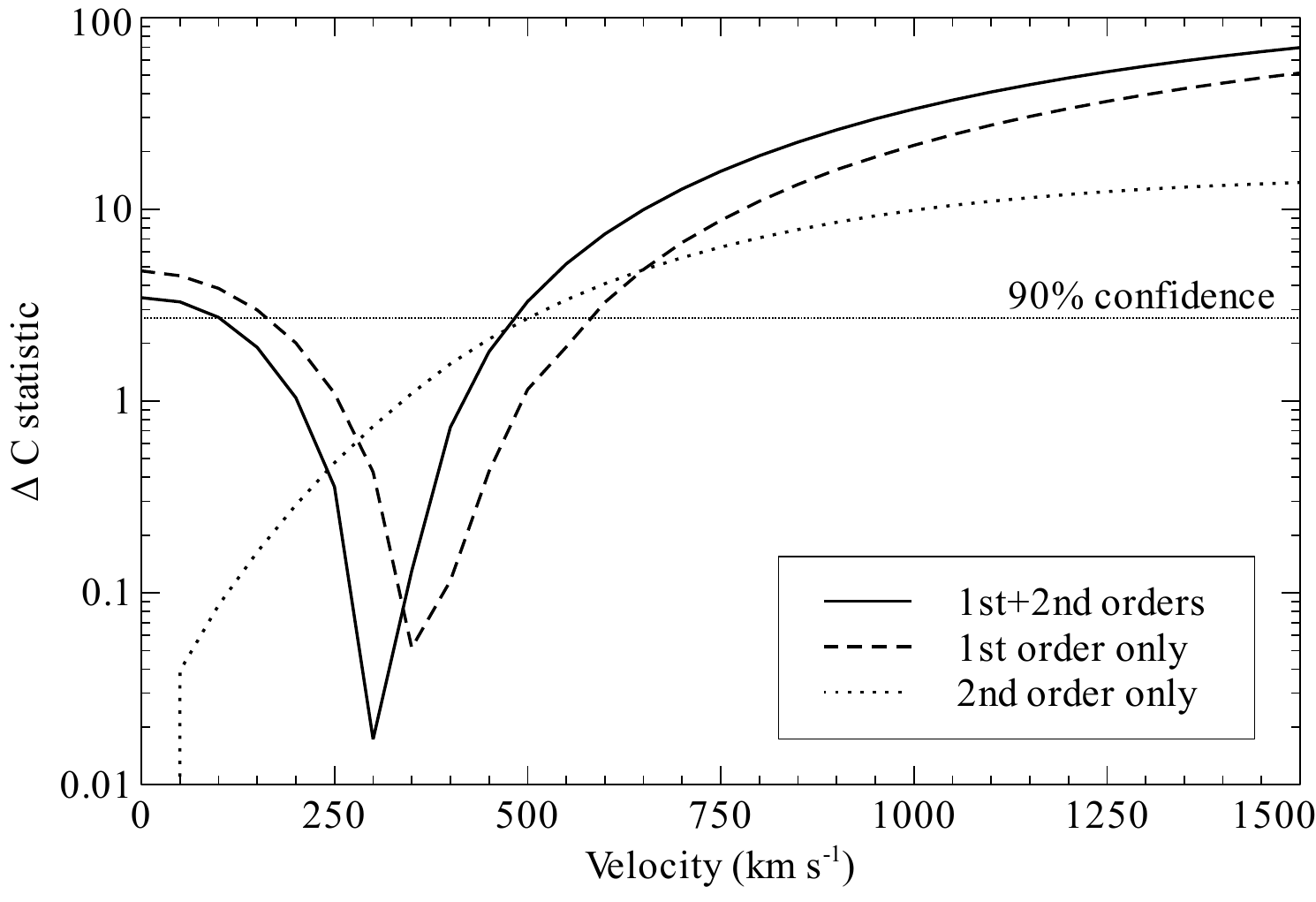}
  \caption{Change in quality of fit as a function of velocity
    broadening for Zw\,3146, fitting both spectral orders or the
    individual spectral orders. The 90 per cent confidence level is
    indicated.}
  \label{fig:zw3146delcstat}
\end{figure}

To demonstrate how we measure our limits, plotted in
Fig.~\ref{fig:zw3146delcstat} is the change in fit quality as a
function of broadening velocity for Zw\,3146. The plot shows that for
this object we are able to very good limits using either of the
spectral orders, or both combined.

For some objects (Abell~262, 2A~0335+096, NGC~4636, Centaurus,
Abell~3581, Abell~1991, Abell~2052 and Abell~S1101), single component
models are poor fits to the spectra. The models do not properly fit
the Fe~\textsc{xvii} emission lines in the spectrum. For those objects
we added a second thermal component to account for the lower
temperature gas in the core of the cluster. This cooler component was
tied to the same metallicities and redshift as the hotter one. We
allowed the cooler component to have a different line broadening. As
this cooler material is concentrated in the cores of these objects it
is likely to have a smaller line width because the spatial broadening
effect is lower. We show the upper limits for the velocity broadening
for the cooler component separately in Table~\ref{tab:vel} and plot
them in Fig.~\ref{fig:velocities}.

We looked for other second thermal components in the remaining sample
of objects by automatically fitting models. We did not find any
objects where the parameters for a second thermal component were
distinct from the first component, or where the new line width was
well constrained.

\subsection{Markov Chain Monte Carlo}
Conventional spectral fitting and error estimation can sometimes
underestimate the likely range of model parameters which can fit
data. An alternative way to determining the model parameter space
probability distribution is to use a Markov Chain Monte Carlo (MCMC)
method.

We applied the MCMC routine built into \textsc{xspec}, which uses the
Metropolis-Hastings algorithm to sample parameter space, constructing
a chain of parameter values. The algorithm starts from a particular
point in parameter space. A new set of parameters is selected by
adding values from a proposal distribution (here a N-dimensional
Gaussian) to the current parameters. If the quality of fit is better
for the new parameters, they are `accepted' and become the next point
in the chain. If they have a worse fit, they are accepted randomly
with a probability which depends on the increase in the fit parameter.
The parameters are otherwise rejected and the next point on the chain
is the existing point.

Such a method explores parameter space around the best fit. The
fraction of chain values in a particular parameter space region should
be related to the likelihood of those model parameters under certain
conditions. These include that the proposal distribution should be big
enough to explore parameter space (it will take too long to converge
if it is too small), but not so large that few points are accepted in
a chain. The length of the chain required to sample parameter space
depends on factors such as the proposal width and how complex the
parameter space is. Unfortunately it is difficult to tell when a chain
has sufficiently converged. We manually examined the results of the
chain over its run to check that it was covering parameter space
randomly and we also checked that the repeat fraction of the chain was
close to the rule-of-thumb value of 0.75, which indicates the proposal
distribution is correct.

We used the \textsc{xspec} MCMC routines to examine the uncertainties
on the the broadening of the emission lines for a number of
objects. We imposed the following priors on the acceptable range of
parameter values:
\begin{enumerate}
\item The redshift of the object should not vary from the NED value by
  more than $1500\kmps$.
\item The Galactic column density to the object should not vary from
  the average values of \cite{Kalberla05} by more than 20 per cent.
\item Elemental abundances should lie between 0.2 and $1\Zsun$.
\end{enumerate}
Although these were not necessary for most objects, for some objects
the chain wandered into very unlikely regions of parameter space and
could not move back. For three objects, Abell 2667, E~1455+2232 and
Kelmola 44, we had to impose a fixed redshift on the chain because of
the data quality.

\begin{figure}
  \includegraphics[width=\columnwidth]{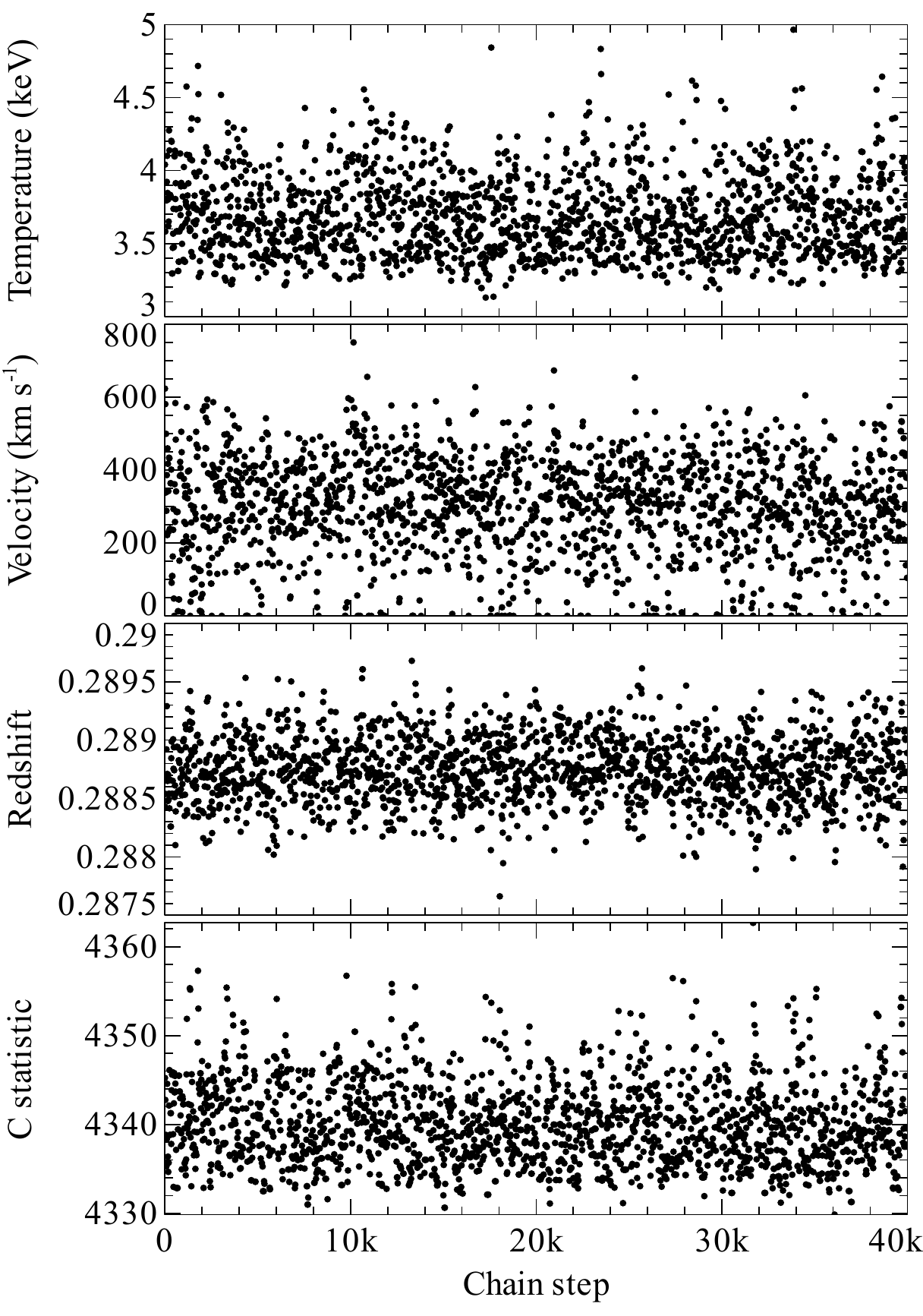}
  \caption{Temperature, velocity, redshift and fit statistic in the
    MCMC chain for Zw\,3146 as a function of the chain step. We plot
    only 1 out of 20 points for clarity.}
  \label{fig:zw3146}
\end{figure}

We used an initial proposal distribution based on the uncertainties of
each parameter using the \textsc{xspec} error command. We started the
chains from the best-fitting values, discarding the first 2000 values
(the burn in period) and using a chain length of 40000 steps. We took
the parameter distribution from this chain and used it to create a
proposal distribution for a second chain with the same length. We
discarded the first chain (used to get a reasonable proposal
distribution) and used the second chain to calculate the results. In
Fig.~\ref{fig:zw3146} are shown the temperature, velocity, redshift
and fit statistic as a function of chain step for Zw\,3146.

In Fig.~\ref{fig:velocities_xtra} are plotted the 90 per cent upper
limit from the chains for the velocity broadening, marginalising over
the other parameters. We get good agreement between our upper limits
and those from conventional spectral fitting for most objects. There
are exceptions to this, including RBS 797, RX\,J2129.6+0005, Abell 2667
and Klemola 44.  These objects appear to have a long tail in the
marginalised probability distribution for the velocity broadening,
often with an inner tighter core. In the case of Klemola 44, the
cluster occupies much of the region used to extract spectral
backgrounds. If a template background is used instead for this object,
the best fitting velocity broadening is $3200^{+950}_{-600}\kmps$,
close to the limit obtained using MCMC.

\subsection{Response line spread function}
Fitting the line shape of bright point-like objects such as Capella
indicates that systematic uncertainties remain in the line spread
function (LSF) of the RGS instruments (den Herder, private
communication).  If the intrinsic spectral resolution is different
from the calibrated value, this would lead to an incorrect
determination of the additional line width caused by the extent of the
source or motions within the object.

We have approached this problem by adjusting the calibration of the
RGS LSF to see the effect on our upper limits. The \textsc{rgsrmfgen}
response generation tool in \textsc{sas} supports the use of an
optional FUDGE header keyword in the FIGURE part of the RGS1 and
RGS2\_LINESPREADFUNC\_0004.CCF calibration files. This parameter is
used to adjust the width of the narrow part of the LSF. We repeated
our analysis using values where this width was increased and decreased
by 10 per cent, using fudge values of 1.1 and 0.9, respectively.

The upper limits derived using these manipulated line spread functions
are also shown on Fig.~\ref{fig:velocities}. The effect of this on
many objects is small, but can be as much as $50\kmps$ for the target
with one of the lowest limits, NGC~4261. A typical increase or
reduction in the limit is around $25\kmps$. Therefore the effect of
the line spread function calibration at the 10 per cent level is
relatively small. Any calibration issues are therefore unlikely to
change the results by more than $100\kmps$.

\section{Differentiating spatial broadening and turbulence}
\begin{figure*}
  \includegraphics[width=0.25\textwidth]{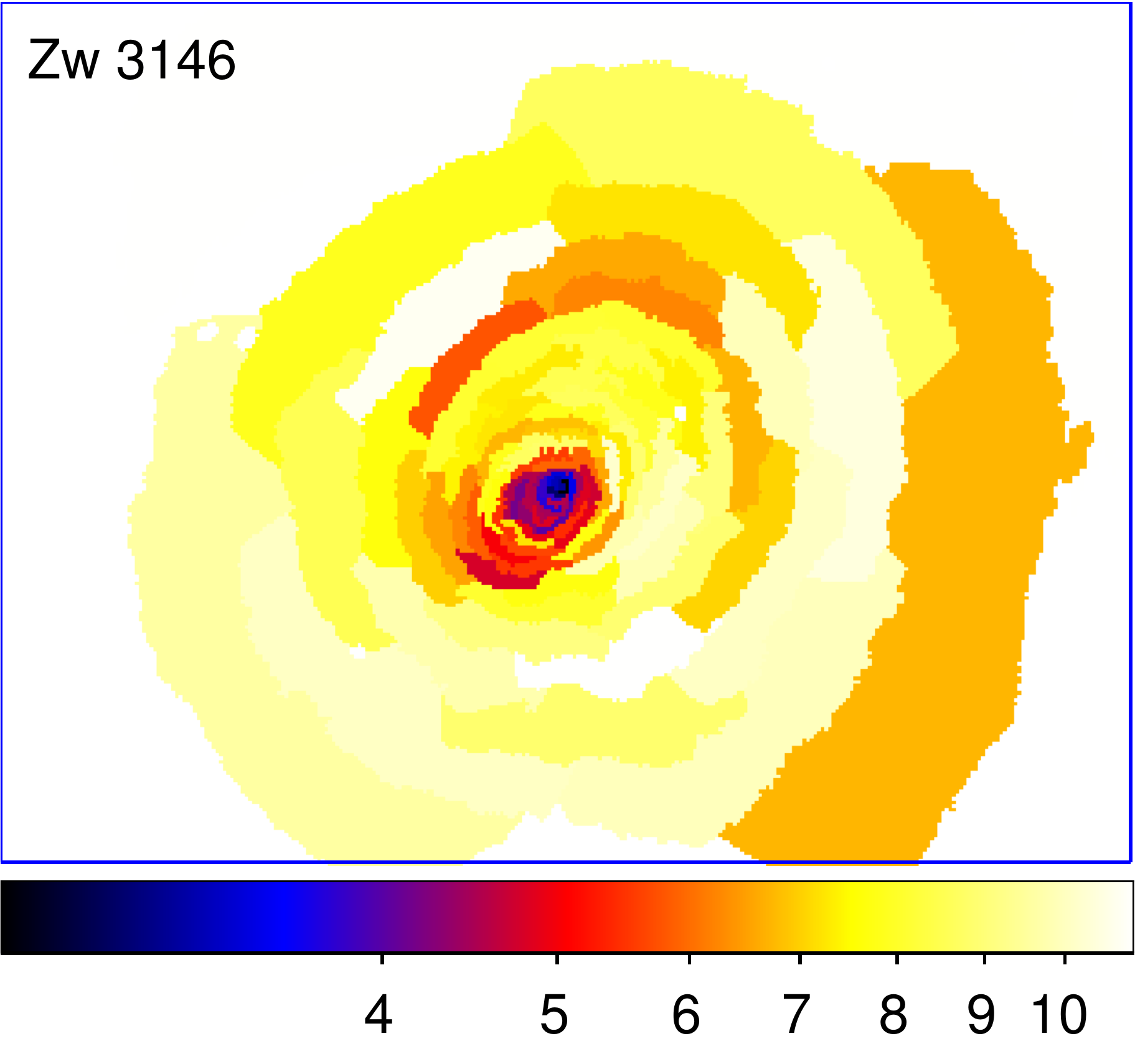}
  \includegraphics[width=0.25\textwidth]{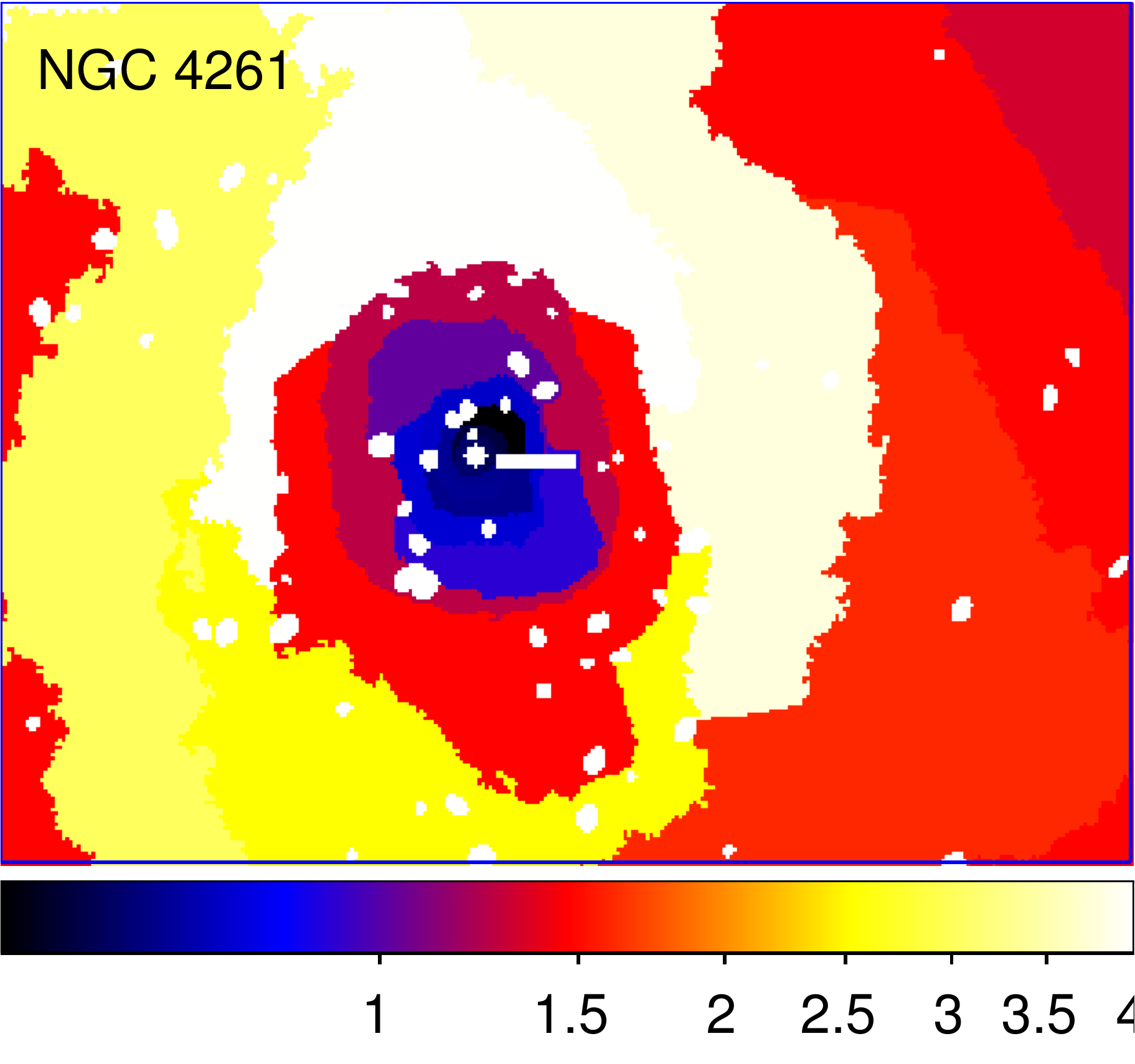}
  \includegraphics[width=0.25\textwidth]{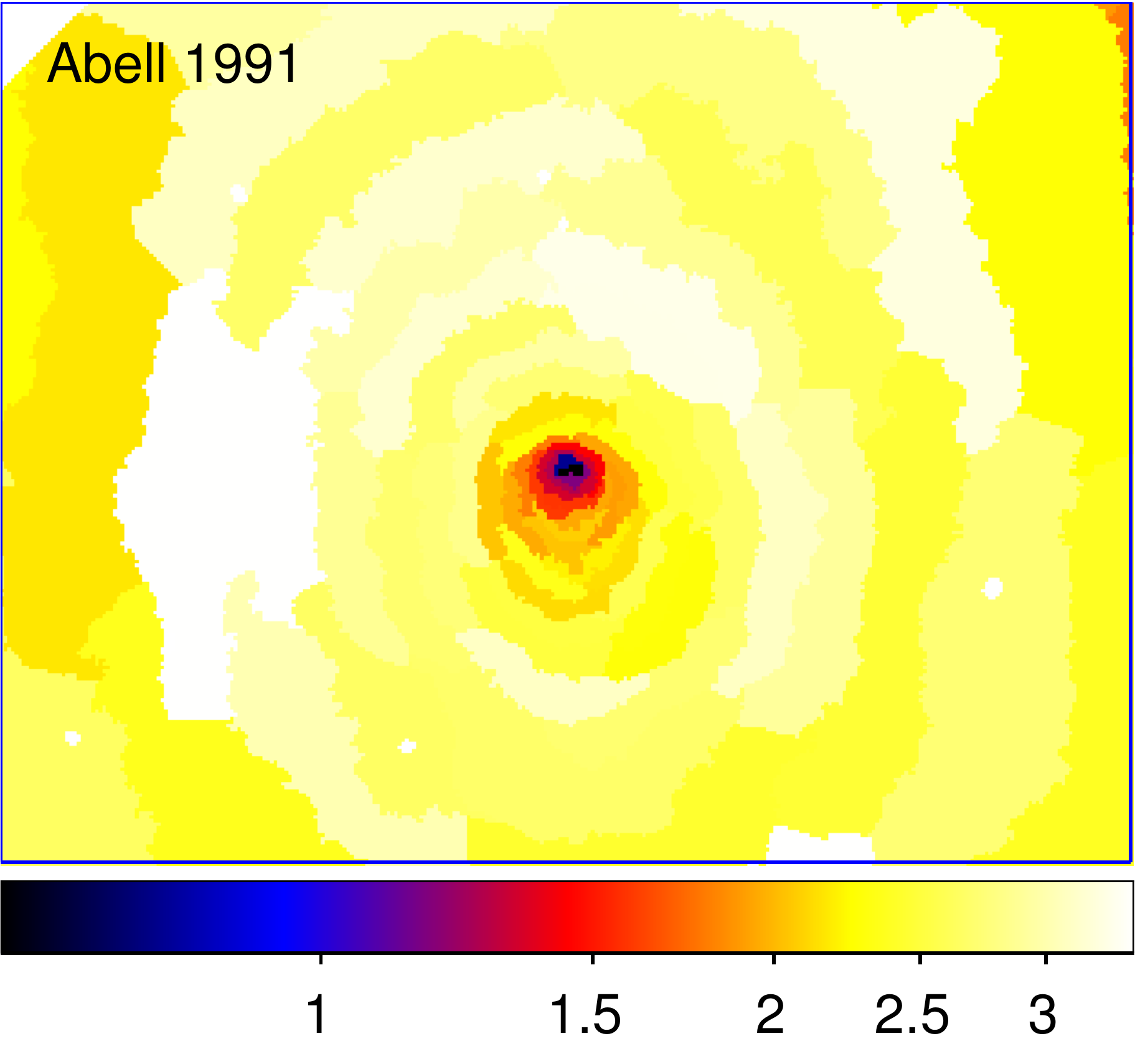}
  \includegraphics[width=0.25\textwidth]{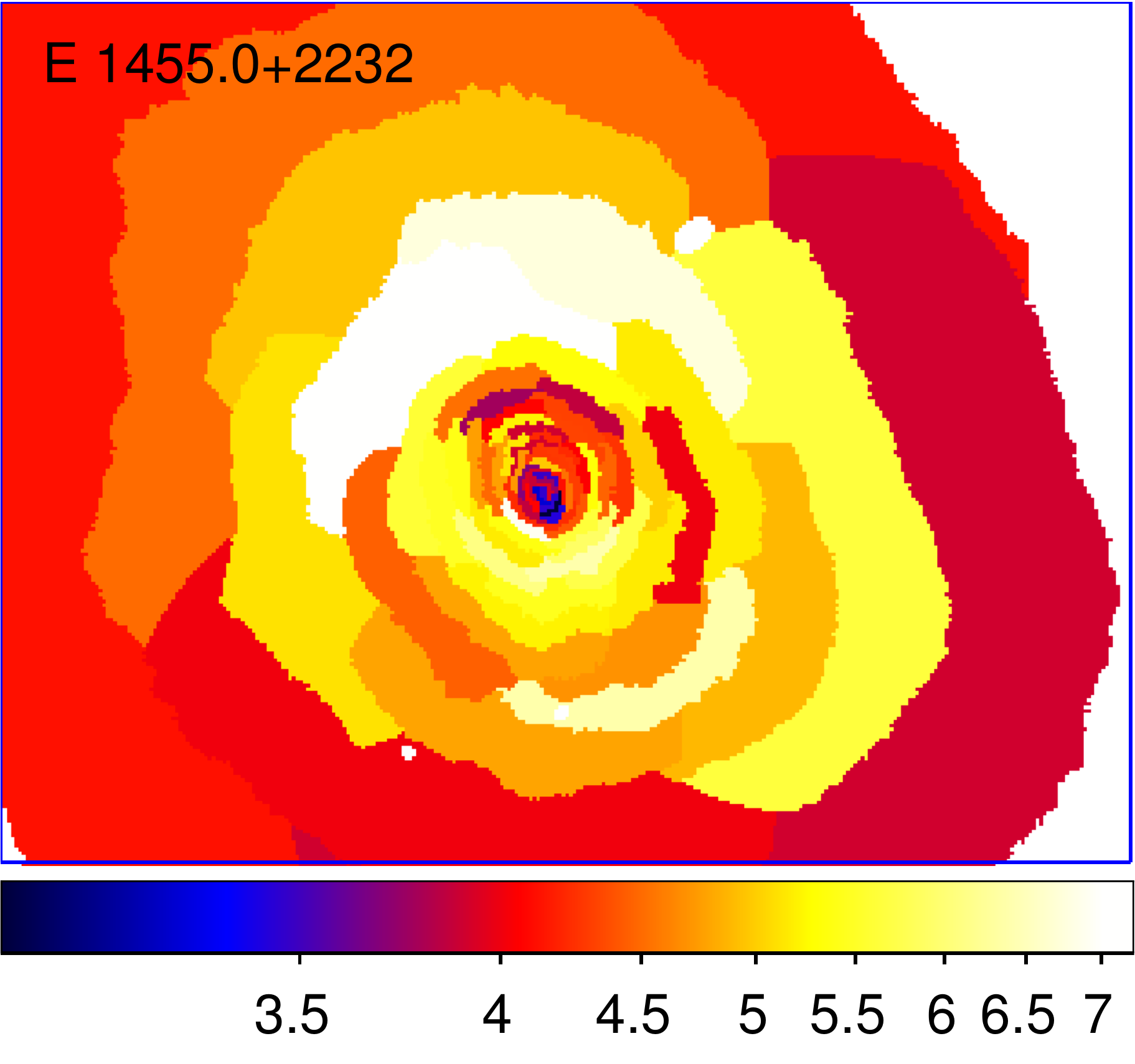}
  \includegraphics[width=0.25\textwidth]{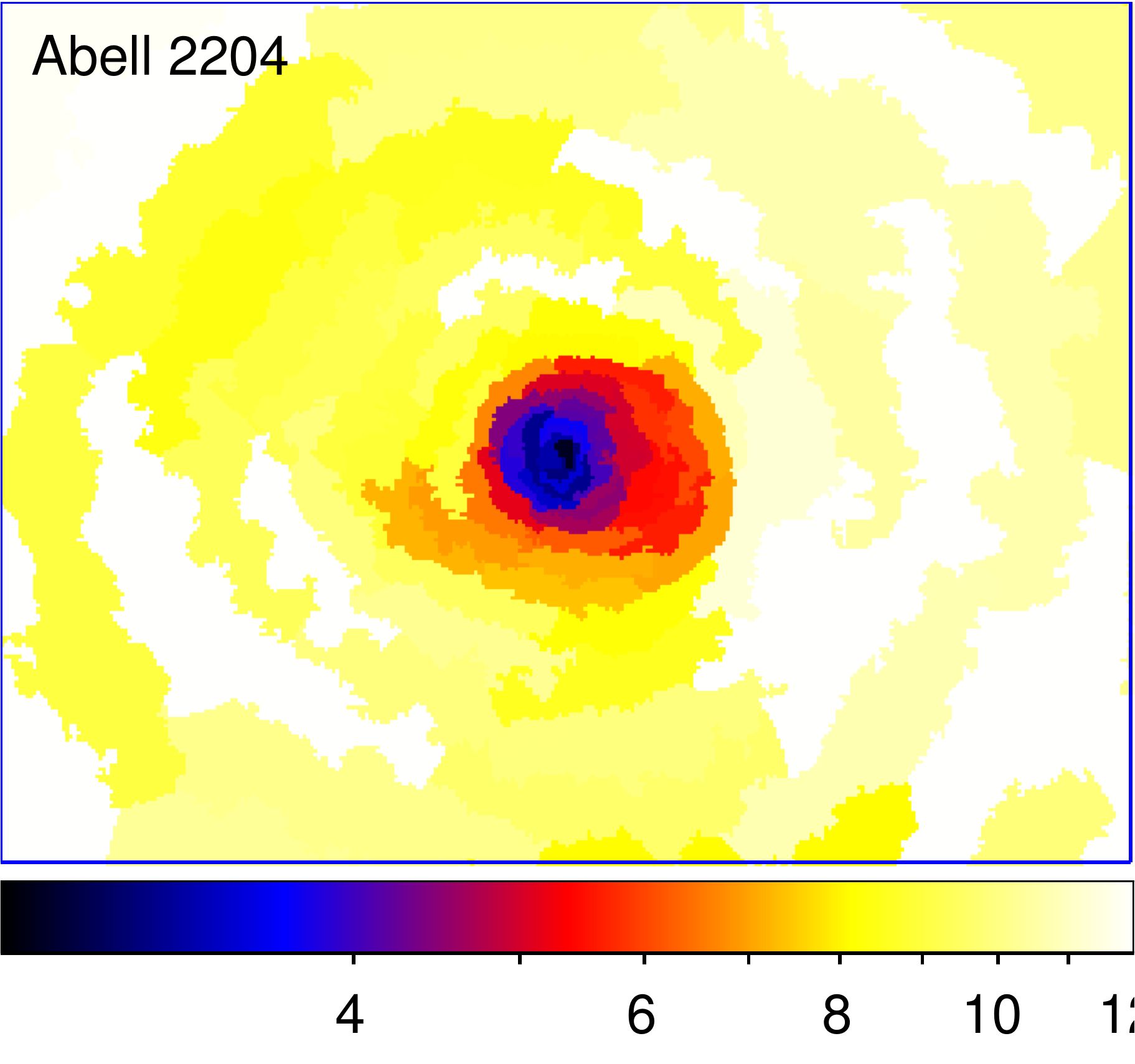}
  \includegraphics[width=0.25\textwidth]{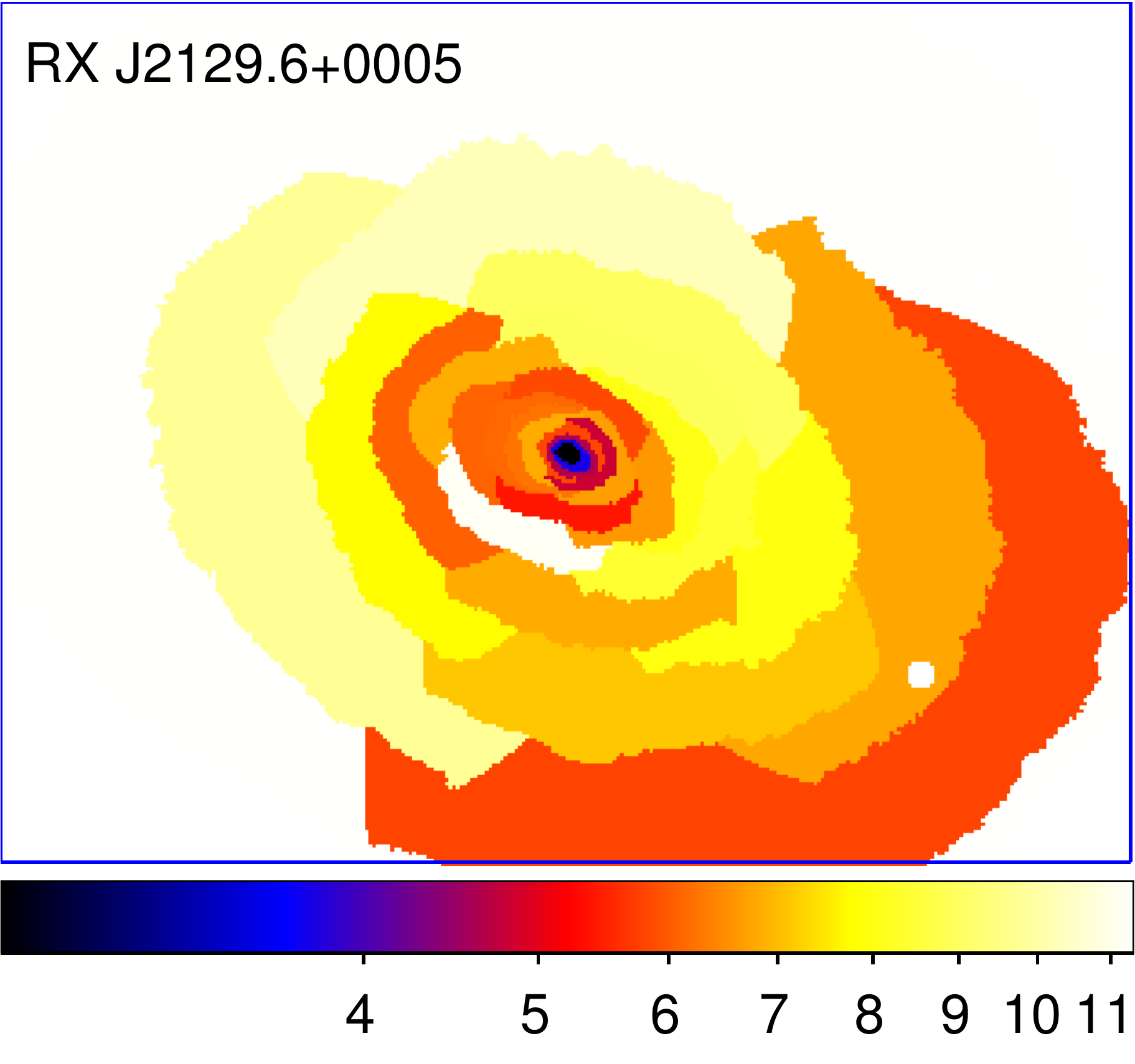}
  \caption{\emph{Chandra} temperature maps for some of the clusters or
    groups with the lowest broadening limits. The scale below each map
    is the temperature in keV. Each region is chosen to contain around
    900 counts or 2000 in the case of Abell 2204. The maps are at the
    same scale on the sky, measuring 5 by 3.9 arcmin. Excluded point
    sources are shown as white regions.}
  \label{fig:tmaps}
\end{figure*}

Those clusters for which we obtained the best limits on broadening
have small compact cores (Fig.~\ref{fig:tmaps}) and so have low
spatial broadening. In order to do better than the limits in Section
\ref{sect:specfitting} we need to differentiate spatial broadening and
intrinsically broad lines, which requires combined spatial and
spectral information about the target.  We need to know which parts of
the RGS spectra come from what dispersion direction angles. This is
done by simulating RGS spectra from \emph{Chandra} spectral maps.

\subsection{Spectral map creation}
Our \emph{Chandra} temperature maps were created by examining the
datasets listed in Table \ref{tab:chandraobs}. The datasets were
reprocessed with the standard \textsc{ciao} tools and very faint event
processing was applied if possible. We used an iterative algorithm to
remove time periods where the count rate was further than three
standard deviations from the median in 200s time bins on the CCDs of
the same type which did not include the object itself (ACIS-0, 1 and 2
for ACIS-I observations, and ACIS-5 for ACIS-S observations). We
excluded point sources, which were initially detected by
\textsc{wavdetect} and confirmed by eye.

The Contour Binning algorithm \citep{SandersBin06} was used to select
regions with similar surface brightness containing a minimum signal to
noise ratio in each bin (listed in Table \ref{tab:chandraobs}) between
0.5 and 7~keV. Spectra were extracted from the event files for each of
these spatial bins. Background spectra were extracted from blank sky
observations, scaled to match the 9 to 12 keV event rate. We created
surface-brightness-weighted response and ancillary response matrices
for each bin. When we examined objects with multiple observations, we
added the spectra for each bin using the same detector, weighting the
response matrices and background spectra according to the relative
foreground exposure times.

The spectra were fitted with an absorbed \textsc{apec} model between
0.5 and 7 keV. The temperature, metallicity and normalization were
free in each fit. Solar relative abundance ratios were assumed, and
absorption was taken from the Galactic values of
\cite{Kalberla05}. Those objects with the best limits have small cool
cores (see Fig.~\ref{fig:tmaps}). For example, Zw\,3146 mostly has only
temperatures below 4 keV in a 4 arcsec radius region in the core of
the cluster. This spatial extent would give a broadening in the Fe-L
complex of around $150-300\kmps$. Our observed limit is $480\kmps$,
which also includes statistical uncertainties.

\subsection{Simulating RGS spectra}
\label{sect:simulating}
To assess whether any of our objects have much broader spectra than
expected from their spatial extent, we simulated RGS spectra without
turbulent broadening using \emph{Chandra} temperature, metallicity and
emission measure maps for a subset of our sample. The \emph{Chandra}
maps were created using the mapping method above.

For each observation combined in the RGS analysis for an object we
deduced which part of the maps were in the RGS extraction region,
using the position of the source, rotation angle of the instrument,
and the PSF extraction region used. For a particular bin in a map, we
calculated the dispersion direction angular distribution of the
emission measure per unit area (assuming it was flat over the bin).  A
response matrix for that bin was then generated using the angdistset
option to the \textsc{rgsrmfgen} tool, for each detector and RGS
order.  In \textsc{xspec} we constructed a spectral model for each bin
using the data from the \emph{Chandra} temperature, metallicity and
emission measure maps, and simulated spectra using this model and the
response matrices for the bin. An absorbed \textsc{apec} model was
used to simulate the spectra, including thermal broadening, but
assuming the same abundance ratio of each element to the Solar
vale. For the simulation an exposure time 100 times greater than the
real exposure time for an observation was used. The resulting spectra
were then added for each detector and for each spectral order. We then
combined the spectra of the same order for all detectors and
observations with \textsc{rgscombine}, as was done for the real data.

These simulated spectra were fit with the same model as we applied to
the data to get the expected line broadening from the spatial extent
in $\kmps$. We did not include instrumental or cosmic backgrounds in
the simulation and did not produce observationally-derived background
spectra in the same way as when extracting the real spectra. The
simulated spectra are therefore the whole spectrum from the extraction
region. We therefore compare the results here using the real spectra
with template blank-sky backgrounds spectra.

There are some weaknesses in this simulation. We do not include our
statistical uncertainties in the \emph{Chandra} maps when simulating
the spectrum. Solar relative abundances are assumed, however most of
the RGS spectral lines are due to Fe and so are the \emph{Chandra}
metallicity measurements, so this is probably not a poor approximation
except for the O~\textsc{viii} line. We do not account for the
cross-dispersion direction change in effective area of the RGS
instruments. However, this is small over the 90 per cent PSF
extraction region we use.  The input maps to the simulation are
projected quantities. The RGS and \emph{Chandra} detectors react
differently to input radiation, so the projected temperatures may well
be different. We do not measure, therefore cannot simulate
true-multiphase material inside a bin.

\begin{figure}
  \includegraphics[width=\columnwidth]{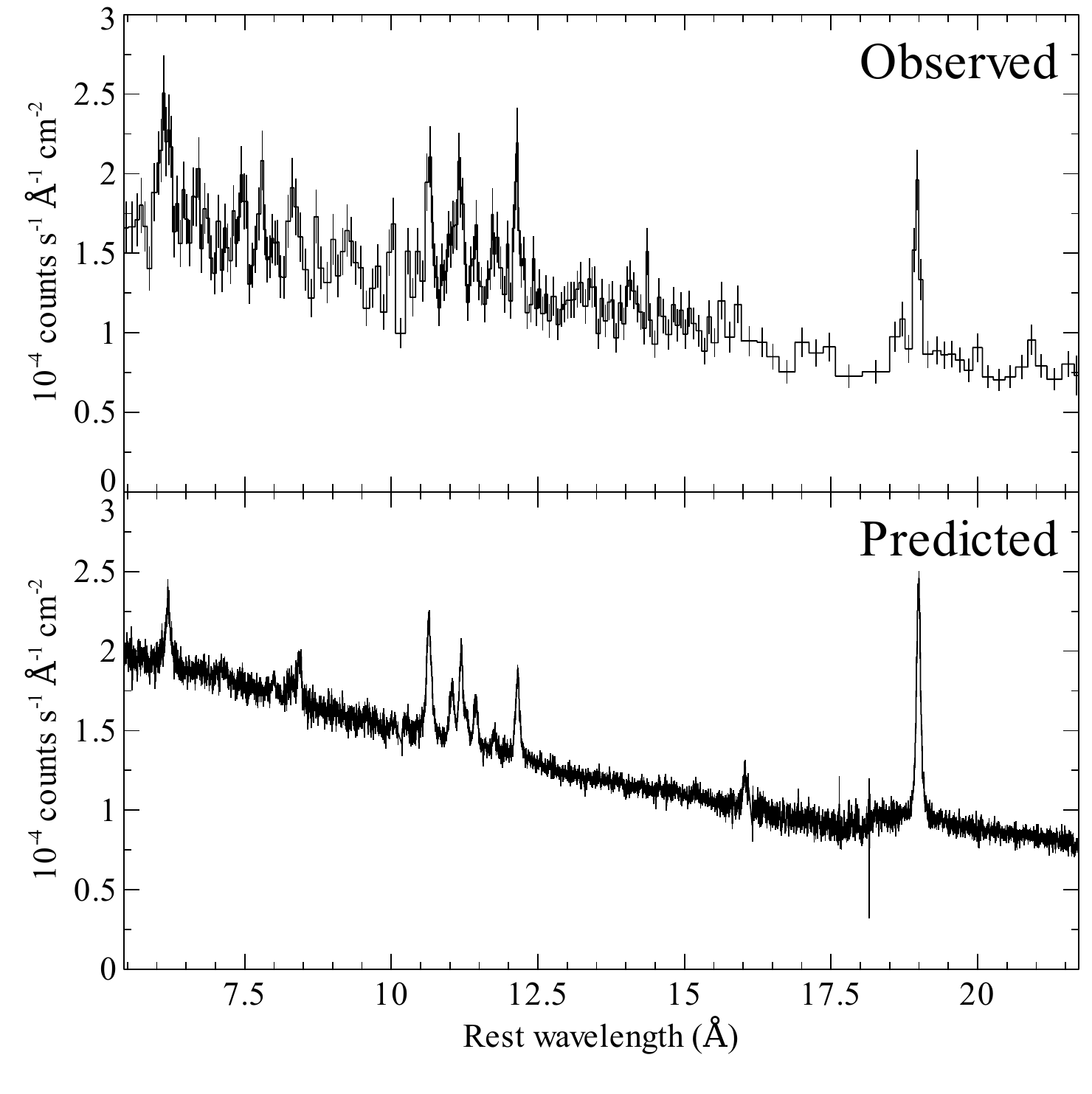}
  \caption{The observed spectra of Zw\,3146 and a spectrum simulated
    using \emph{Chandra} spectral mapping.The observed spectrum was
    rebinned to have a minimum signal to noise ratio of 10.}
  \label{fig:zw3146spec}
\end{figure}

The simulation does relatively well when tested with cool core compact
targets. For some hotter targets the simulations have too much of the
continuum emission at short wavelengths. In the case of Zw\,3146
(Fig.~\ref{fig:zw3146spec}), the agreement between our simulations and
the observed spectra at low wavelengths is good. For Abell 1795, the
simulation contains 20 per cent more flux at {7\AA} than
observed. There is reasonable agreement of the line strengths between
the real and simulated spectra, however.

Table \ref{tab:linewidths} shows a comparison between the real line
widths and measured simulation line widths. We have not included
targets where we know there evidence for multiphase gas in the core of
the object (e.g. Centaurus, Abell~262 and Abell~3581), as our present
\emph{Chandra} modelling only fits the spectrum with a single
component at each point. For some objects we found there was a
discrepancy between the determination of the line widths using
standard statistics and using a MCMC analysis. For these we show the
MCMC results which tend to be larger (indicated in the table).

There are two objects we have modelled which show evidence for
turbulent broadening: Klemola~44 and RX\,J1347.5-1145.  If we take the
best fitting model for the Klemola~44 data using a template
background, forcing this to have a broadening equal to the that
obtained from the simulated spectrum, then we get a C statistic 4.4
higher than the best fitting model.  This corresponds to a 3.6 per
cent probability of the extra broadening of $1300\kmps$ happening by
chance. The probability derived using MCMC with a template background,
marginalising over the other parameters in the chain, is much lower at
0.7 per cent.

In the case of RX\,J1347.5-1145, the evidence for an excess broadening
is weaker, with a change probability of the increase of $840\kmps$
seen by chance of 6 per cent, as calculated from the change in fit
statistic. The marginalised probability from the MCMC chains is
actually much lower at 0.6 per cent.  It is a hot cluster, with a
lowest temperature of 5.9~keV in our \emph{Chandra} temperature map,
which makes the measurement of emission lines fairly difficult.

Two of the objects show upper limits which are negative (Abell 2597
and NGC 5044). This is clearly unphysical, but we would expect a
similar number of negative values given the sample size and that we
are using 90 per cent upper limits, if turbulence is intrinsically
small. The measured line widths for these objects are close to the
predicted values. We note that two-thirds of the objects have
predicted line widths greater than their measured values. A $50\kmps$
systematic increase in the measured values would give an even balance
and remove the two negative upper limits.

\subsection{Zw 3146}
Presented in this paper are deep new observations of the luminous
galaxy cluster Zw 3146. This adds an additional 176~ks onto the 55~ks
of data in the archive. The spectrum for this cluster is shown in
Fig.~\ref{fig:spectralt500} and Fig.~\ref{fig:zw3146spec}. Our limit
on the broadening of this cluster is $480\kmps$, which is dominated by
the spatial extent of the source. If we assume that the expected
broadening from the spatial extent from our modelling is correct
(Section \ref{sect:simulating}), then we are able to convert this to
a 90 per cent upper limit on the turbulence of $155\kmps$.

\begin{figure}
  \includegraphics[width=\columnwidth]{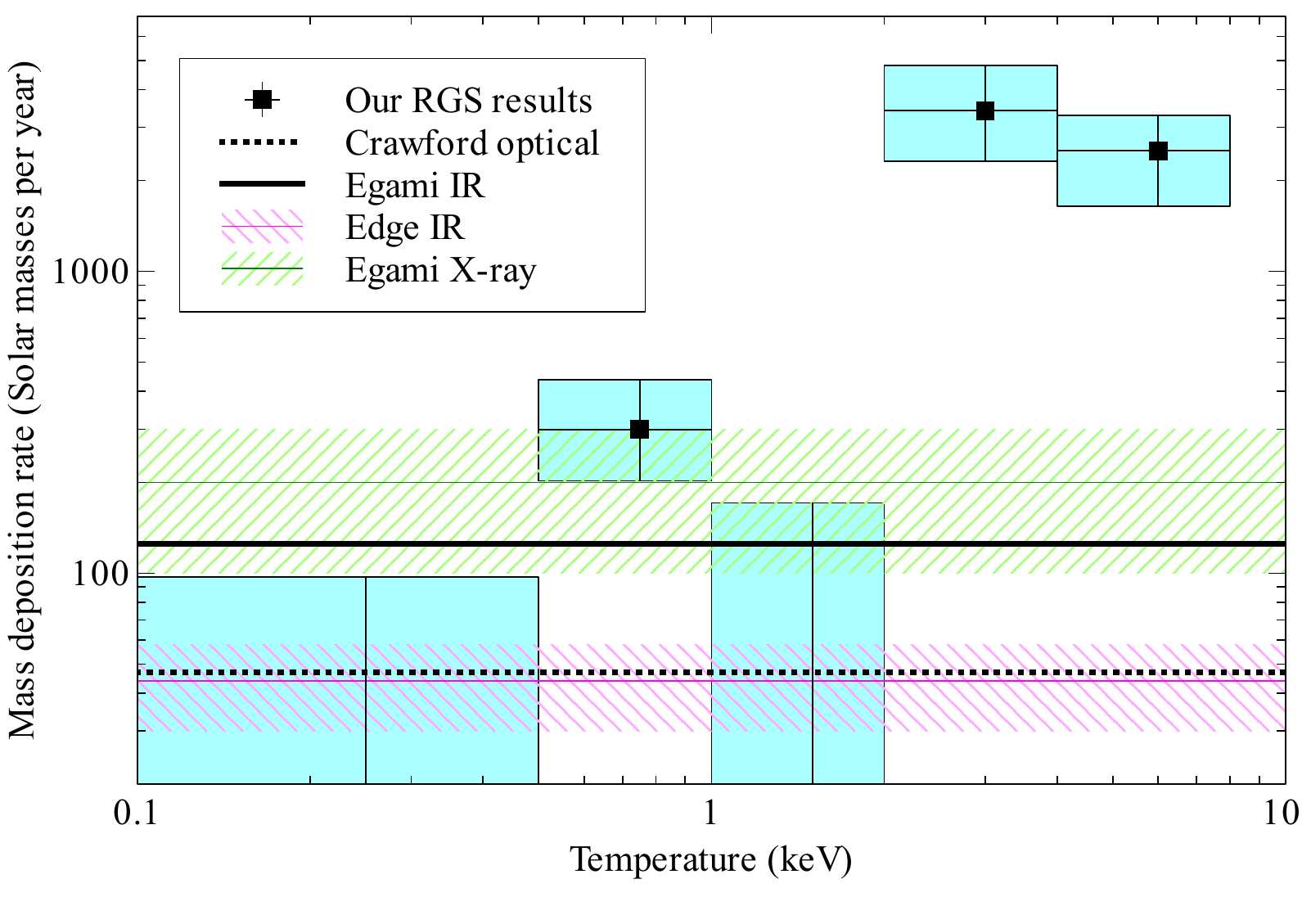}
  \caption{Mass deposition rate as a function of temperature for
    Zw\,3146. The boxes show the 1$\sigma$ uncertainties on mass
    deposition rate and ranges or temperature for each component in
    the model. Also shown are the mass deposition rates from
    \protect\cite{Egami06}, \protect\cite{Edge10} and
    \protect\cite{Crawford99}.}
  \label{fig:zw3146_mdot}
\end{figure}

As this is a deep observation we have determined what the distribution
on the amount of gas as a function is. As for Abell 1835
\citep{Sanders10_A1835}, we fitted a model made up of gas cooling
through ranges of temperature. We use ranges 8 to 4, 4 to 2, 2 to 1, 1
to 0.5 and 0.5 to 0.08~keV. The amount of gas in each bin is
parameterized as the mass deposition rate through each temperature
range (the number of Solar masses of material which could be cooling
radiatively according to the cooling function). We use the best
fitting line width and redshift from the previous observation in this
analysis and allow the same metals to have free abundances.

The mass deposition rates as a function of temperature are shown in
Fig.~\ref{fig:zw3146_mdot}. Also plotted are the \cite{Egami06}
results for cooling rates from \emph{Chandra} and \emph{Spitzer},
\cite{Edge10} results using \emph{Herschel} and the optical star
formation rate from \cite{Crawford99}. Our results are consistent with
the observed star formation rates in the optical and infrared
wavebands coming from cooling X-ray gas. In the lowest temperature bin
our result is below the Egami IR measurement, but this is not
statistically significant.

Like in Abell 1835, we find that there is mild evidence for a
component at around 0.7~keV in the spectrum, which is equivalent in
normalization to a cooling rate of around $300\Msunpyr$.

\section{Discussion}
\subsection{Turbulence}
\begin{figure}
  \includegraphics[width=\columnwidth]{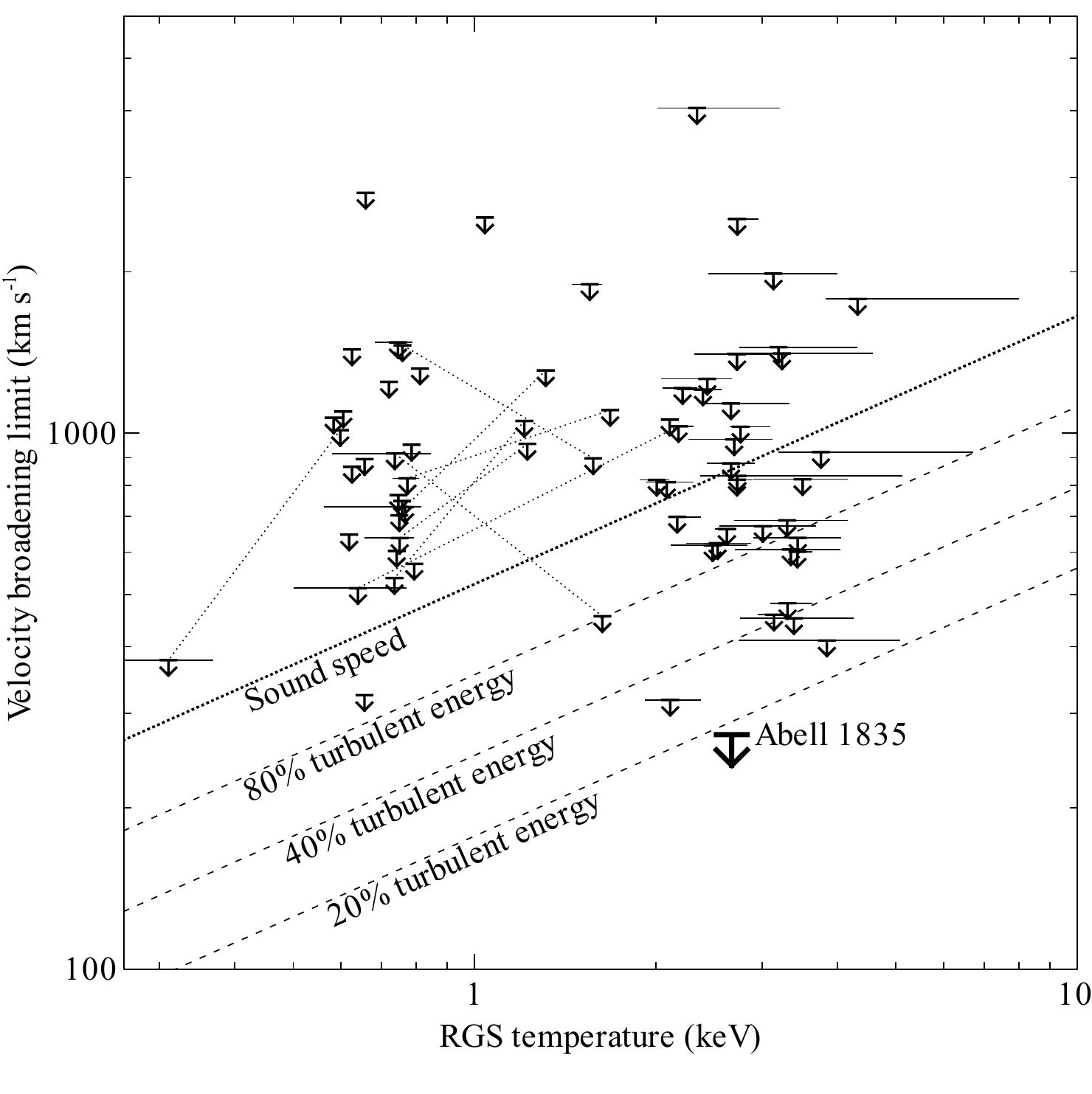}
  \caption{Limits on velocity broadening as a function of RGS-measured
    temperature. Also plotted are the sound speed as a function of
    temperature, and what fraction of the thermal energy density the
    velocity broadening would represent. The points connected by
    dotted lines show the results using one and two temperature
    components for the same objects. Abell~1835 is analysed in
    \protect \cite{Sanders10_A1835}.}
  \label{fig:velocities_kT}
\end{figure}

Plotted in Fig.~\ref{fig:velocities_kT} are the upper limits for the
sample of objects against RGS temperature. Note that the plotted
temperature is measured from the same region as the velocity
limit. The central region is likely to be a cool core, and is not the
temperature of the bulk of the cluster. We also plot the sound speed
as a function of temperature. Shown in addition are lines of constant
fractions of turbulent to thermal energy density, calculated using
equation 11 from \cite{Werner09},
\begin{equation}
 \frac{\varepsilon_\mathrm{turb}}{\varepsilon_\mathrm{therm}} =
 \frac{V_\mathrm{los}^2}{kT} \: \mu m_\mathrm{p},
\end{equation}
where $V_\mathrm{los}$ is the measured line-of-sight velocity, $\mu$
is the mean particle mass, $m_\mathrm{p}$ is the proton mass and $kT$
is the RGS temperature.

Any trends in this plot should be ignored as the points are upper
limits and it is dominated by selection effects. We are able to obtain
limits on the broadening less than the sound speed in many objects,
and less than half the sound speed in some objects, but unfortunately
none of our new limits are better than our previous Abell 1835
limit. The optimal temperature to measure limits relative to the sound
speed with the RGS appears to be $\sim 3$~keV.

If we make the step of assuming that the spatial contribution to the line
widths of the RGS spectra from our simulations are correct (Table
\ref{tab:linewidths}), we can get better limits for some objects by
subtracting the component from the spatial size. Our general result is
that there is little evidence for any extra broadening in our spectra
due to turbulence.  We find typical limits for galaxy clusters with
small error bars on the measured results, such as Zw\,3146, Abell~496,
Abell~1795, Abell~2204 and HCG~62 of only a few hundred \kmps.

\begin{figure}
  \includegraphics[width=\columnwidth]{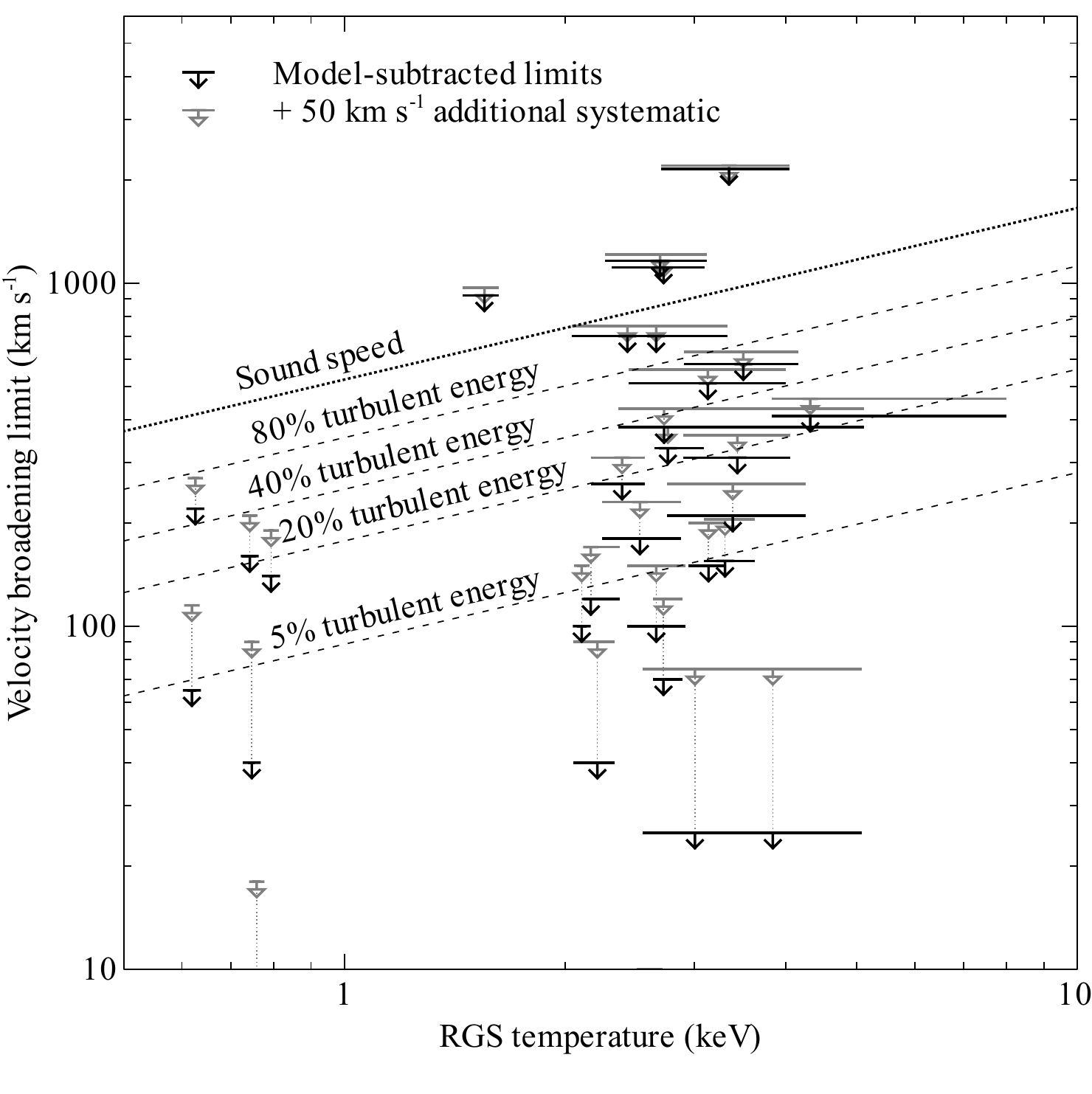}
  \caption{Limits on velocity broadening after subtracting the
    contribution from the source spatial extent using modelling. These
    results should be treated with caution as they are dependent on
    the modelling being correct. We also show the results after adding
    an additional $50 \kmps$ possible systematic uncertainty from the
    LSF calibration.}
  \label{fig:velocities_kT_submodel}
\end{figure}

Fig.~\ref{fig:velocities_kT_submodel} shows the upper limits we obtain
after subtracting the contribution from the spatial extent of the
source as a function of temperature. Even after including a $50\kmps$
additional systematic uncertainty, we find $\sim 15$ sources which
have less than 20 per cent of the thermal energy in turbulence.

We find two objects which may have broadening of the emission lines by
gas motions. One is Klemola~44, which appears to be turbulently
broadened by around $1300\kmps$. The disturbed morphology, lack of a
strong central peak and lack of cool core in this object (as seen in
our \emph{Chandra} spectral mapping) indicate that it is not
relaxed. The cluster hosts a $\sim 60 \kpc$ long extended radio
structure $\sim 40 \kpc$ from the cluster core \citep{SleeRoy98}. This
may be a relic indicative of a cluster merger, however, it is smaller
and has a steeper radio spectrum than is typical in merger relics, but
indicates some sort of major disturbance. In addition the galaxies in
the core of the cluster have a much lower velocity dispersion
($127\kmps$, \citealt{Green90}) than the cluster as a whole ($847
\kmps$). The core is also offset by around $500\kmps$ in mean velocity
relative to the surrounding cluster. These observations imply that the
cluster has a disturbed velocity structure, agreeing with our findings
of line broadening.

The other cluster which hints at turbulent motions at the $840\kmps$
level is RX\,J1347.5-1145.  This object appears to have recently
undergone a merger, as seen through X-ray substructure and hot $\sim
20\keV$ material in a region only a few 10s of arcsec from the core
\citep{Allen02,GittiSchindler04,Miranda08,Bradac08,Ota08}. These
substructures have also been in Sunyaev-Zel'dovich observations
\citep{Komatsu01,Mason10}. Recent lensing and velocity measurements
point towards a merger in the plane of the sky \citep{Lu10}. The cool
region in the core of this cluster is only a few arcsec wide, so very
good limits would be possible were it not for the large amounts of
continuum emission from surrounding gas. Our marginal detection of
turbulent broadening in this object is consistent with the merger
scenario. Deeper observations of this object could improve the
evidence for turbulence.

In order to make better limits or measurements of the turbulence in
clusters with \emph{XMM}, observations need to be targeted on compact
bright objects with cool cores at moderate redshift. The flux from
emission lines needs to be maximized and the source size minimized, to
obtain the best upper limits. In these objects we may become limited
by the calibration of the LSF.

The launch of \emph{ASTRO-H} in the future should provide an X-ray
calorimeter with energy resolution of $\sim 7$~eV over the 0.3 to
12~keV band. This will revolutionize the measurement of velocities in
the intracluster medium. The X-ray microcalorimeter spectrometer on
\emph{IXO} in the further future is planned to provide 2.5~eV
resolution imaging spectroscopy over the inner $2\times2$ arcmin of
objects, and 10~eV resolution out to $5.4\times5.4$ arcmin.

\subsection{Redshifts}
\begin{figure}
  \includegraphics[width=\columnwidth]{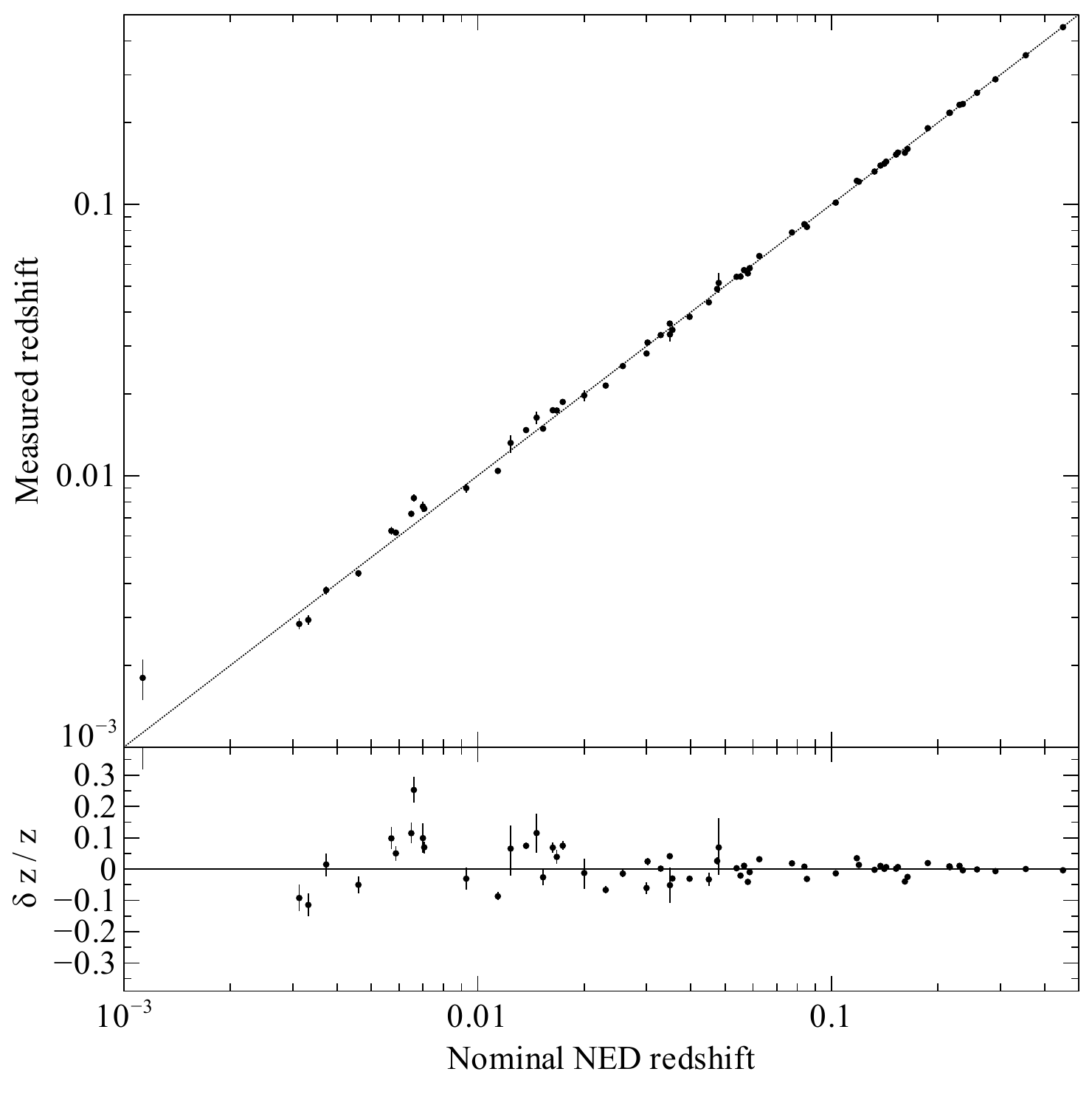}
  \caption{Measurements of redshift compared to the nominal values
    taken from NED, using single thermal component models. The bottom
    panel shows the fractional deviation of our measured results from
    the NED values.}
  \label{fig:redshifts}
\end{figure}

\begin{figure}
  \includegraphics[width=\columnwidth]{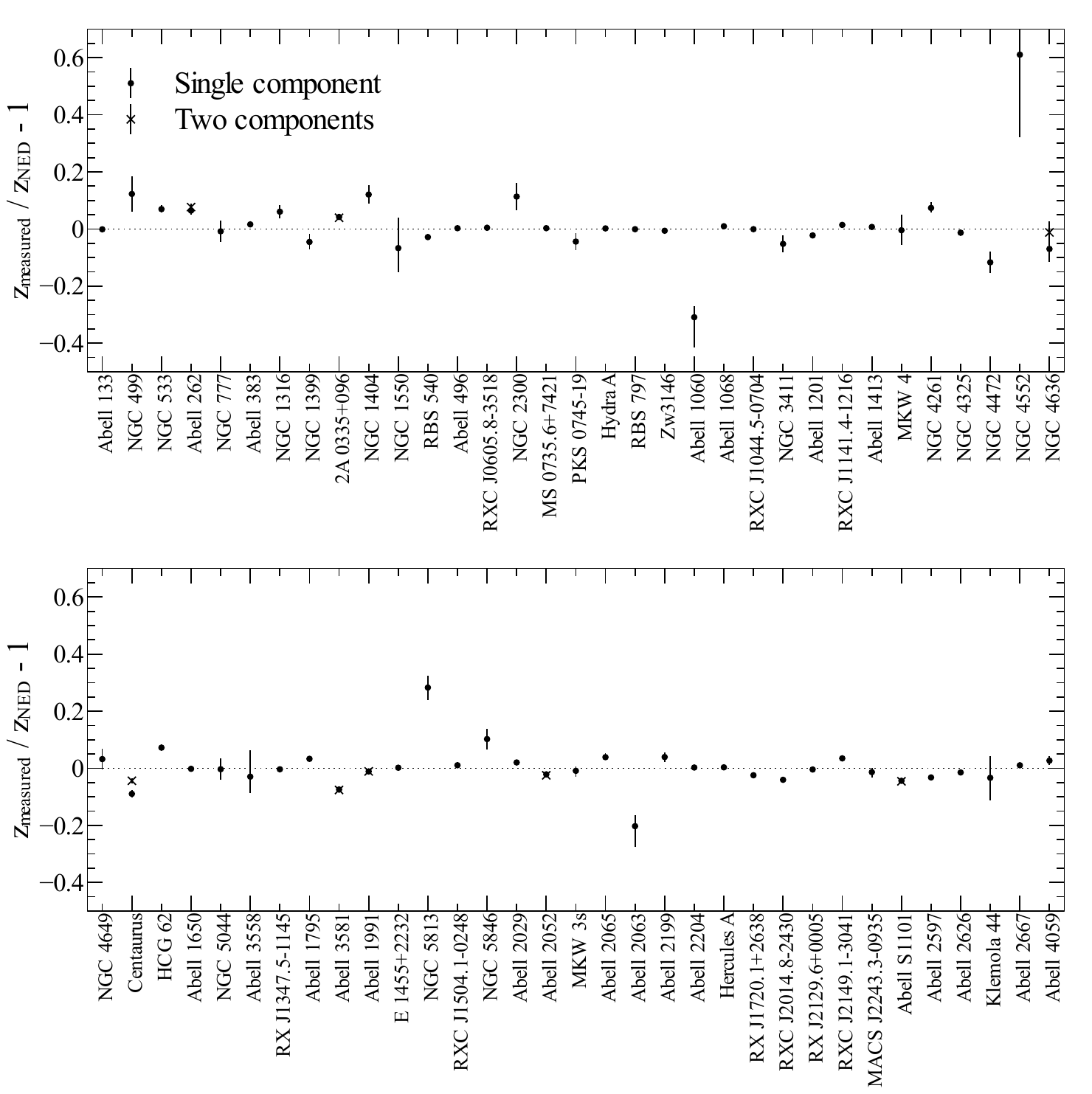}
  \caption{Fractional difference between our measured redshifts and
    those taken from the NED database. We show the single and two
    thermal component results as appropriate.}
  \label{fig:redshiftdev}
\end{figure}

In Fig.~\ref{fig:redshifts} our measured redshifts are plotted against
the value from the NED database and the fractional difference as a
function of redshift. We show in Fig.~\ref{fig:redshiftdev} the
fractional difference between the two values for each object. Although
the agreements between our values and those from NED are good in many
objects, there are a few targets for which there are a significant
discrepancy. This may be due to errors in the analysis, such as
problems identifying the position of the centre of the cooler X-ray
emitting material if the object is particularly
extended. Alternatively they could identify a real velocity difference
between the optical and X-ray emission lines.  There are no obvious
systematic offsets between the NED redshifts and our results.

\begin{figure}
  \includegraphics[width=\columnwidth]{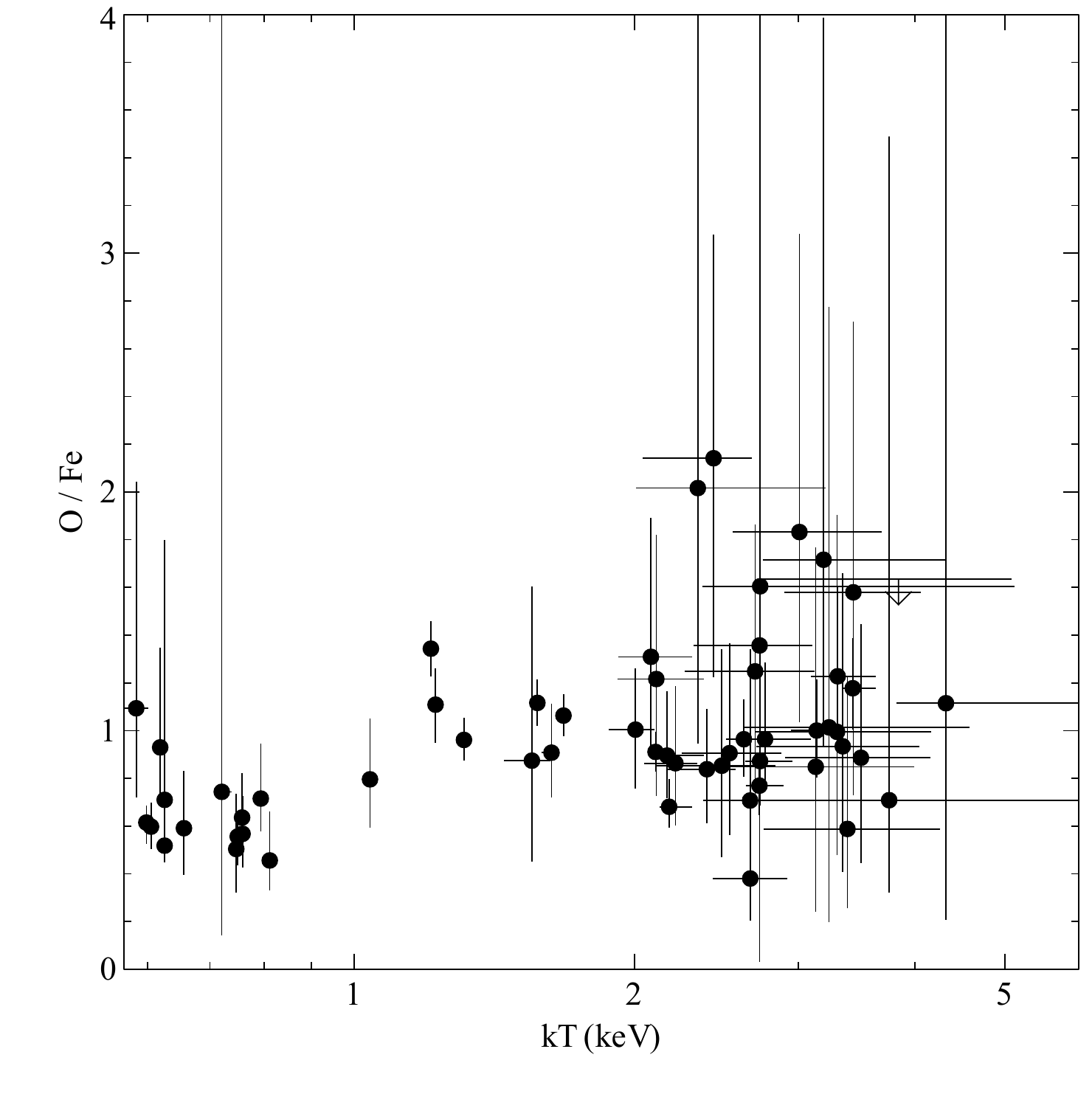}
  \caption{O to Fe ratio, in Solar units, plotted against temperature
    for the sample. Note that some cooler objects are constrained to
    have Solar Fe metallicity.}
  \label{fig:oferatio}
\end{figure}

\subsection{Metallicities}
We are able to measure the metallicity of several elements in our
objects (Table \ref{tab:param}). The uncertainty on the metallicities
depend quite strongly on temperature because the presence of emission
lines is a function of temperature. The absolute metallicities depend
on how well the continuum is modelled in the spectral fit. As the
extraction regions (a slit) will contain the outer parts of the
objects and we only use a single temperature model for most of the
fits, this can give extra continuum and lower
metallicities. Metallicity ratios are more robust than absolute
metallicities, especially if there is extra continuum. We plot the
O/Fe abundance ratio as a function of temperature for our sample in
Fig.~\ref{fig:oferatio}. We find the Fe/O ratio is similar between the
objects and are consistent with Solar. There may be some evidence for
lower O/Fe in cooler systems, although our sample has strong selection
biases and we are likely to be examining larger spatial regions in
nearby groups than the more distant and hotter clusters.

\section{Conclusions}
We examine \emph{XMM-Newton} observations of a sample of galaxy
clusters, galaxy groups and elliptical galaxies chosen to have a
compact core. By fitting thermal models to their RGS spectra, we place
limits on the velocity broadening of their spectra. If we ignore the
fact that the objects are extended, we can place conservative upper
limits of less than $500 \kmps$ of broadening for five targets. Half
of the objects examined have velocity limits of less than $700\kmps$.
After subtracting the spatial component of the broadening by
modelling, we find two objects with significant velocity
broadening. Klemola 44 which is broadened by around $1500\kmps$ and
RX\,J1347.5-1145 by $800\kmps$.  In addition, if our modelling of the
contribution to the broadening of the spatial extent of the objects is
correct, we obtain limits on turbulence for Zw\,3146, Abell~496,
Abell~1795, Abell~2204 and HCG~62 of less than $200\kmps$. Including a
$50\kmps$ additional calibration systematic uncertainty, we find 15
sources with less than 20 per cent of the thermal energy density in
turbulence.

\section*{Acknowledgements}
ACF thanks the Royal Society for support. We also thank J.-W. den
Herder and G.~Pratt for useful discussions.

This research has made use of the NASA/IPAC Extragalactic Database
(NED) which is operated by the Jet Propulsion Laboratory, California
Institute of Technology, under contract with the National Aeronautics
and Space Administration.

\bibliographystyle{mnras}
\small
\bibliography{refs}

\onecolumn
\setlength{\LTcapwidth}{\textwidth}

\small
\begin{longtable}{llllp{3cm}lll}
  \caption{List of objects, position used for extraction, nominal
    redshift, \emph{XMM-Newton} datasets analysed, sum of the RGS1 and
    RGS2 cleaned exposure times,
    background-subtracted count rate, and background count rate.}\\
  \label{tab:obj}
  Cluster & RA (J2000) & Dec (J2000) & Redshift & Datasets & Exposure (ks) & Fg. Rate (\ps) & Bg. Rate (\ps) \\ \hline
         Abell 133 & $01:02:41.76$ & $-21:52:49.8$ &   0.0566 &  0144310101 & 67.5 & 0.065 & 0.026 \\
         NGC 499 & $01:23:11.65$ & $+33:27:38.6$ &  0.01467 &  0501280101 & 107.8 & 0.009 & 0.003 \\
         NGC 533 & $01:25:31.39$ & $+01:45:33.1$ &  0.01739 &  0109860101 & 68.3 & 0.019 & 0.007 \\
       Abell 262 & $01:52:46.18$ & $+36:09:10.2$ &   0.0163 &  0109980101 0504780101/\-0201 & 264.4 & 0.041 & 0.019 \\
         NGC 777 & $02:00:14.92$ & $+31:25:45.8$ & 0.016728 &  0203610301 0304160301 & 119.0 & 0.025 & 0.014 \\
       Abell 383 & $02:48:03.36$ & $-03:31:44.7$ &   0.1871 &  0084230501 & 64.6 & 0.037 & 0.007 \\
        NGC 1316 & $03:22:41.72$ & $-37:12:28.3$ &  0.00587 &  0302780101 0502070201 & 343.2 & 0.016 & 0.006 \\
        NGC 1399 & $03:38:29.08$ & $-35:27:02.0$ &   0.0046 &  0400620101 & 259.2 & 0.041 & 0.015 \\
     2A 0335+096 & $03:38:41.10$ & $+09:58:00.7$ &   0.0349 &  0147800201 & 267.7 & 0.152 & 0.038 \\
        NGC 1404 & $03:38:51.74$ & $-35:35:40.4$ &  0.00649 &  0304940101 & 74.1 & 0.048 & 0.015 \\
        NGC 1550 & $04:19:37.91$ & $+02:24:35.8$ & 0.012389 &  0152150101 & 61.0 & 0.043 & 0.018 \\
         RBS 540 & $04:25:51.27$ & $-08:33:36.4$ &   0.0397 &  0300210401 & 82.3 & 0.057 & 0.017 \\
       Abell 496 & $04:33:37.86$ & $-13:15:41.9$ &   0.0329 &  0135120201 0506260301/\-0401 & 339.1 & 0.103 & 0.038 \\
RXC J0605.8-3518 & $06:05:53.98$ & $-35:18:09.2$ &    0.141 &  0201901001 & 73.7 & 0.049 & 0.027 \\
        NGC 2300 & $07:32:19.70$ & $+85:42:33.3$ &    0.007 &  0022340201 & 104.9 & 0.010 & 0.005 \\
  MS 0735.6+7421 & $07:41:44.24$ & $+74:14:38.2$ &    0.216 &  0303950101 & 143.0 & 0.025 & 0.010 \\
     PKS 0745-19 & $07:47:31.19$ & $-19:17:39.9$ &   0.1028 &  0105870101 & 43.3 & 0.099 & 0.023 \\
         Hydra A & $09:18:05.99$ & $-12:05:43.9$ &   0.0539 &  0504260101 & 244.1 & 0.139 & 0.035 \\
         RBS 797 & $09:47:12.69$ & $+76:23:13.4$ &    0.354 &  0502940301 & 62.9 & 0.050 & 0.020 \\
          Zw3146 & $10:23:39.65$ & $+04:11:11.2$ &   0.2906 &  0108670101 0605540201/\-0301 & 426.2 & 0.064 & 0.009 \\
      Abell 1068 & $10:40:44.52$ & $+39:57:10.3$ &   0.1375 &  0147630101 & 45.5 & 0.074 & 0.010 \\
RXC J1044.5-0704 & $10:44:32.85$ & $-07:04:08.7$ &   0.1323 &  0201901501 & 57.8 & 0.058 & 0.011 \\
        NGC 3411 & $10:50:26.08$ & $-12:50:42.5$ &   0.0153 &  0146510301 & 71.0 & 0.052 & 0.018 \\
RXC J1141.4-1216 & $11:41:24.43$ & $-12:16:39.7$ &   0.1195 &  0201901601 & 72.2 & 0.032 & 0.019 \\
      Abell 1413 & $11:55:17.89$ & $+23:24:21.8$ &   0.1427 &  0112230501 0502690101/\-0201 & 344.5 & 0.036 & 0.017 \\
           MKW 4 & $12:04:27.17$ & $+01:53:44.8$ &     0.02 &  0093060101/\-0301 & 38.2 & 0.031 & 0.014 \\
        NGC 4261 & $12:19:23.23$ & $+05:49:29.6$ &  0.00706 &  0502120101 & 245.0 & 0.013 & 0.007 \\
        NGC 4325 & $12:23:06.62$ & $+10:37:15.6$ &  0.02571 &  0108860101 & 42.6 & 0.049 & 0.011 \\
        NGC 4472 & $12:29:46.78$ & $+08:00:02.9$ &  0.00332 &  0200130101 & 203.7 & 0.051 & 0.019 \\
        NGC 4552 & $12:35:39.82$ & $+12:33:22.8$ &  0.00113 &  0141570101 & 86.8 & 0.019 & 0.013 \\
        NGC 4636 & $12:42:49.92$ & $+02:41:15.3$ &  0.00313 &  0111190701 & 124.4 & 0.063 & 0.019 \\
        NGC 4649 & $12:43:39.95$ & $+11:33:09.7$ &  0.00373 &  0502160101 & 163.9 & 0.047 & 0.011 \\
       Centaurus & $12:48:48.95$ & $-41:18:45.0$ &   0.0114 &  0046340101 0406200101 & 339.8 & 0.137 & 0.050 \\
          HCG 62 & $12:53:05.71$ & $-09:12:15.2$ &   0.0137 &  0112270701 0504780501/\-0601 & 354.7 & 0.028 & 0.011 \\
      Abell 1650 & $12:58:41.50$ & $-01:45:44.3$ &   0.0838 &  0093200101 & 83.0 & 0.051 & 0.019 \\
        NGC 5044 & $13:15:23.95$ & $-16:23:07.5$ &  0.00928 &  0037950101 & 46.4 & 0.093 & 0.032 \\
      Abell 3558 & $13:27:56.85$ & $-31:29:44.1$ &    0.048 &  0107260101 & 86.1 & 0.035 & 0.020 \\
 RX J1347.5-1145 & $13:47:30.59$ & $-11:45:10.1$ &    0.451 &  0112960101 & 73.8 & 0.054 & 0.009 \\
      Abell 1795 & $13:48:52.36$ & $+26:35:37.3$ &   0.0625 &  0097820101 & 128.7 & 0.186 & 0.085 \\
      Abell 3581 & $14:07:29.78$ & $-27:01:05.9$ &    0.023 &  0205990101 0504780301/\-0401 & 375.4 & 0.064 & 0.022 \\
      Abell 1991 & $14:54:31.62$ & $+18:38:41.5$ &   0.0587 &  0145020101 & 83.7 & 0.049 & 0.014 \\
     E 1455+2232 & $14:57:15.09$ & $+22:20:32.5$ &   0.2578 &  0108670201 & 58.2 & 0.042 & 0.025 \\
        NGC 5813 & $15:01:11.26$ & $+01:42:07.2$ &   0.0066 &  0302460101 & 73.1 & 0.059 & 0.018 \\
RXC J1504.1-0248 & $15:04:07.42$ & $-02:48:15.7$ &   0.2153 &  0401040101 & 77.8 & 0.155 & 0.014 \\
        NGC 5846 & $15:06:29.35$ & $+01:36:17.8$ &   0.0057 &  0021540101/\-0501 & 80.9 & 0.034 & 0.013 \\
      Abell 2029 & $15:10:56.14$ & $+05:44:40.5$ &   0.0773 &  0111270201 0551780201/\-0301/\-0401/\-0501 & 385.7 & 0.167 & 0.045 \\
      Abell 2052 & $15:16:44.51$ & $+07:01:17.0$ &   0.0355 &  0109920101 0401520301/\-0501/\-0601/\-0801/\-0901/\-1101/\-1201/\-1601/\-1701 & 417.1 & 0.090 & 0.040 \\
          MKW 3s & $15:21:51.88$ & $+07:42:31.8$ &    0.045 &  0109930101 & 105.9 & 0.105 & 0.039 \\
      Abell 2063 & $15:23:04.85$ & $+08:36:20.2$ & 0.034937 &  0200120401 0550360101 & 99.6 & 0.022 & 0.015 \\
      Abell 2199 & $16:28:38.27$ & $+39:33:03.2$ &   0.0302 &  0008030201/\-0301/\-0601 & 90.1 & 0.113 & 0.062 \\
      Abell 2204 & $16:32:46.96$ & $+05:34:31.1$ &   0.1522 &  0112230301 0306490101/\-0201/\-0301/\-0401 & 229.5 & 0.146 & 0.026 \\
      Hercules A & $16:51:08.18$ & $+04:59:32.6$ &    0.154 &  0401730101/\-0201/\-0301 & 250.0 & 0.022 & 0.009 \\
 RX J1720.1+2638 & $17:20:09.94$ & $+26:37:29.1$ &    0.164 &  0500670201/\-0301/\-0401 & 141.0 & 0.065 & 0.013 \\
RXC J2014.8-2430 & $20:14:51.69$ & $-24:30:20.5$ &   0.1612 &  0201902201 & 53.2 & 0.078 & 0.012 \\
 RX J2129.6+0005 & $21:29:39.94$ & $+00:05:18.8$ &    0.235 &  0093030201 & 113.0 & 0.039 & 0.011 \\
RXC J2149.1-3041 & $21:49:07.72$ & $-30:42:04.7$ &   0.1179 &  0201902601 & 53.6 & 0.032 & 0.008 \\
     Abell S1101 & $23:13:58.76$ & $-42:43:34.7$ &    0.058 &  0147800101 & 250.9 & 0.130 & 0.038 \\
      Abell 2597 & $23:25:19.78$ & $-12:07:27.6$ &   0.0852 &  0108460201 0147330101 & 275.5 & 0.109 & 0.032 \\
      Abell 2626 & $23:36:30.52$ & $+21:08:48.2$ &   0.0553 &  0083150201 0148310101 & 112.9 & 0.032 & 0.012 \\
      Klemola 44 & $23:47:43.18$ & $-28:08:34.8$ &     0.03 &  0204460101 & 59.6 & 0.058 & 0.028 \\
      Abell 2667 & $23:51:39.40$ & $-26:05:02.8$ &     0.23 &  0148990101 & 61.7 & 0.064 & 0.012 \\
      Abell 4059 & $23:57:01.00$ & $-34:45:32.8$ &   0.0475 &  0109950101/\-0201 & 108.0 & 0.046 & 0.022 \\

  \hline
\end{longtable}

\begin{longtable}{lcccccccc}
  \caption{Spectral fitting results for temperature, column density
    and metallicities. Uncertainties on values are at the $1\sigma$
    level. If the Fe metallicity is 1, this means it was fixed at this
    value in the spectral fit. Column density (n$_\mathrm{H}$) is in
    units of $10^{20} \psqcm$. Metallicities
    are relative to Solar.}\\
  \label{tab:param}
  Cluster & kT (keV) & n$_\mathrm{H}$ & O & Ne & Mg & Si & Fe & Ni \\ \hline
         Abell 133 & $2.39 \pm 0.20$ & $<2$ & $0.70 \pm 0.14$ & $0.71^{+0.31}_{-0.26}$ & $0.81 \pm 0.32$ & $2^{+1}_{-1}$ & $0.83 \pm 0.17$ & $1.70 \pm 0.96$ \\
         NGC 499 & $0.659 \pm 0.012$ & $5 \pm 4$ & $0.52 \pm 0.12$ & $<0.28$ & $1.69 \pm 0.37$ & $<2$ & 1 & $2 \pm 1$ \\
         NGC 533 & $0.743 \pm 0.017$ & $7^{+4}_{-3}$ & $0.70^{+0.18}_{-0.14}$ & $1.16 \pm 0.34$ & $0.87 \pm 0.36$ & $<0.55$ & 1 & $4 \pm 1$ \\
       Abell 262 & $1.223 \pm 0.020$ & $8.48 \pm 0.89$ & $0.313 \pm 0.031$ & $0.331 \pm 0.070$ & $0.327 \pm 0.073$ & $<0.65$ & $0.282 \pm 0.029$ & $1.52 \pm 0.20$ \\
         NGC 777 & $0.721^{+0.018}_{-0.011}$ & $<13$ & $0.68^{+17}_{-0.36}$ & $1.26^{+12}_{-0.77}$ & $1^{+2}_{-1}$ & $<890$ & $0.92^{+8}_{-0.56}$ & $4^{+200}_{-3}$ \\
       Abell 383 & $2.66^{+0.67}_{-0.29}$ & $<2$ & $0.30^{+0.19}_{-0.12}$ & $0.43^{+0.43}_{-0.23}$ & $0.48^{+0.54}_{-0.35}$ & $0.73^{+0.47}_{-0.26}$ & $0.42^{+0.26}_{-0.11}$ & $<0.45$ \\
        NGC 1316 & $0.6191 \pm 0.0082$ & $2^{+2}_{-1}$ & $0.251^{+0.072}_{-0.042}$ & $0.199^{+0.10}_{-0.055}$ & $0.219^{+0.12}_{-0.071}$ & $12^{+8}_{-4}$ & $0.270^{+0.094}_{-0.048}$ & $0.60^{+0.29}_{-0.17}$ \\
        NGC 1399 & $0.7939 \pm 0.0068$ & $9 \pm 2$ & $0.242^{+0.046}_{-0.031}$ & $0.351^{+0.12}_{-0.069}$ & $0.388^{+0.13}_{-0.079}$ & $<5$ & $0.339^{+0.088}_{-0.049}$ & $2.00^{+0.58}_{-0.35}$ \\
     2A 0335+096 & $1.573 \pm 0.017$ & $25.44 \pm 0.39$ & $0.311 \pm 0.023$ & $0.307 \pm 0.034$ & $0.287 \pm 0.031$ & $0.17 \pm 0.15$ & $0.279 \pm 0.013$ & $0.371 \pm 0.093$ \\
        NGC 1404 & $0.6052 \pm 0.0079$ & $<0.18$ & $0.238 \pm 0.025$ & $0.324 \pm 0.067$ & $0.217 \pm 0.083$ & $<0.56$ & $0.398 \pm 0.050$ & $0.80 \pm 0.24$ \\
        NGC 1550 & $1.040 \pm 0.017$ & $13 \pm 2$ & $0.294 \pm 0.077$ & $0.30 \pm 0.18$ & $0.68 \pm 0.21$ & $<1$ & $0.369^{+0.061}_{-0.025}$ & $1.92 \pm 0.61$ \\
         RBS 540 & $2.168^{+0.21}_{-0.050}$ & $11 \pm 1$ & $0.485^{+0.10}_{-0.079}$ & $0.42^{+0.19}_{-0.15}$ & $0.69^{+0.22}_{-0.18}$ & $2 \pm 1$ & $0.541^{+0.12}_{-0.062}$ & $0.64 \pm 0.52$ \\
       Abell 496 & $2.107 \pm 0.031$ & $6.37 \pm 0.37$ & $0.409 \pm 0.029$ & $0.576 \pm 0.070$ & $0.505 \pm 0.075$ & $<0.38$ & $0.449 \pm 0.026$ & $0.32 \pm 0.20$ \\
RXC J0605.8-3518 & $3.30^{+0.86}_{-0.61}$ & $5 \pm 1$ & $0.61^{+0.34}_{-0.23}$ & $0.87^{+0.58}_{-0.37}$ & $0.96^{+0.67}_{-0.44}$ & $0.50^{+0.51}_{-0.37}$ & $0.62^{+0.45}_{-0.22}$ & $1^{+2}_{-1}$ \\
        NGC 2300 & $0.657 \pm 0.015$ & $<1$ & $0.47 \pm 0.12$ & $0.81 \pm 0.35$ & $1.06 \pm 0.74$ & $<1$ & 1 & $3 \pm 1$ \\
  MS 0735.6+7421 & $3.01^{+0.68}_{-0.46}$ & $2 \pm 1$ & $0.68^{+0.27}_{-0.20}$ & $0.42^{+0.25}_{-0.20}$ & $0.58^{+0.43}_{-0.32}$ & $0.50^{+0.29}_{-0.22}$ & $0.37^{+0.21}_{-0.12}$ & $1^{+1}_{-1}$ \\
     PKS 0745-19 & $2.73 \pm 0.39$ & $39^{+2}_{-1}$ & $<0.45$ & $0.28 \pm 0.13$ & $0.18 \pm 0.13$ & $<0.45$ & $0.158 \pm 0.055$ & $<0.97$ \\
         Hydra A & $2.724^{+0.17}_{-0.091}$ & $3.72 \pm 0.37$ & $0.235 \pm 0.030$ & $0.322 \pm 0.066$ & $0.131 \pm 0.079$ & $<0.43$ & $0.305 \pm 0.033$ & $<0.41$ \\
         RBS 797 & $3.39^{+0.87}_{-0.63}$ & $5 \pm 1$ & $0.19^{+0.14}_{-0.10}$ & $0.34^{+0.31}_{-0.21}$ & $0.55^{+0.54}_{-0.40}$ & $0.32^{+0.29}_{-0.20}$ & $0.316^{+0.25}_{-0.057}$ & $<0.48$ \\
          Zw3146 & $3.31^{+0.32}_{-0.21}$ & $3.44 \pm 0.49$ & $0.225^{+0.040}_{-0.034}$ & $0.388^{+0.092}_{-0.074}$ & $0.17 \pm 0.11$ & $0.380 \pm 0.075$ & $0.184^{+0.046}_{-0.030}$ & $0.39 \pm 0.34$ \\
      Abell 1068 & $2.53^{+0.35}_{-0.28}$ & $<2$ & $0.271^{+0.10}_{-0.078}$ & $0.38^{+0.19}_{-0.14}$ & $0.47^{+0.24}_{-0.20}$ & $0.31 \pm 0.20$ & $0.299^{+0.098}_{-0.074}$ & $1.10^{+0.89}_{-0.75}$ \\
RXC J1044.5-0704 & $2.48 \pm 0.36$ & $1 \pm 1$ & $0.211^{+0.098}_{-0.073}$ & $0.31^{+0.18}_{-0.13}$ & $0.42^{+0.24}_{-0.19}$ & $<0.39$ & $0.247 \pm 0.076$ & $<0.34$ \\
        NGC 3411 & $0.8117 \pm 0.0085$ & $5 \pm 2$ & $0.227^{+0.064}_{-0.050}$ & $0.19 \pm 0.12$ & $0.79^{+0.36}_{-0.23}$ & $<4$ & $0.497^{+0.18}_{-0.082}$ & $2.33^{+0.98}_{-0.54}$ \\
RXC J1141.4-1216 & $2.08^{+0.22}_{-0.16}$ & $2 \pm 1$ & $0.51^{+0.15}_{-0.13}$ & $0.65^{+0.32}_{-0.22}$ & $0.35^{+0.35}_{-0.28}$ & $<0.31$ & $0.388^{+0.13}_{-0.081}$ & $<0.53$ \\
      Abell 1413 & $3.23^{+1}_{-0.62}$ & $3 \pm 1$ & $0.099^{+0.094}_{-0.064}$ & $0.23^{+0.26}_{-0.12}$ & $0.34^{+0.32}_{-0.16}$ & $0.32^{+0.30}_{-0.16}$ & $0.098^{+0.14}_{-0.047}$ & $<2$ \\
           MKW 4 & $1.552 \pm 0.090$ & $<8$ & $0.84^{+0.52}_{-0.29}$ & $2.25^{+1}_{-0.87}$ & $<1$ & $<17$ & $0.96^{+0.55}_{-0.33}$ & $<2$ \\
        NGC 4261 & $0.6565 \pm 0.0091$ & $<1$ & $0.262 \pm 0.065$ & $0.50^{+0.28}_{-0.10}$ & $0.33 \pm 0.17$ & $<12$ & $0.442^{+0.14}_{-0.094}$ & $1.01^{+0.62}_{-0.30}$ \\
        NGC 4325 & $0.786 \pm 0.011$ & $<3$ & $0.472 \pm 0.091$ & $1.42 \pm 0.32$ & $1.18 \pm 0.29$ & $<0.62$ & 1 & $2 \pm 1$ \\
        NGC 4472 & $0.7499 \pm 0.0065$ & $<1$ & $0.308 \pm 0.047$ & $0.67 \pm 0.13$ & $0.56 \pm 0.12$ & $10 \pm 3$ & $0.554 \pm 0.071$ & $2.46 \pm 0.43$ \\
        NGC 4552 & $0.584 \pm 0.017$ & $<4$ & $0.346^{+0.27}_{-0.084}$ & $0.29^{+0.26}_{-0.12}$ & $0.83^{+0.70}_{-0.28}$ & $<5$ & $0.316^{+0.13}_{-0.076}$ & $1.09^{+0.97}_{-0.44}$ \\
        NGC 4636 & $0.5983 \pm 0.0056$ & $<0.15$ & $0.272 \pm 0.029$ & $0.253 \pm 0.046$ & $0.345 \pm 0.061$ & $<0.48$ & $0.442 \pm 0.034$ & $1.01^{+0.23}_{-0.16}$ \\
        NGC 4649 & $0.7580^{+0.011}_{-0.0086}$ & $<3$ & $0.386^{+0.085}_{-0.068}$ & $0.57^{+0.17}_{-0.14}$ & $0.81 \pm 0.21$ & $7^{+4}_{-4}$ & $0.606^{+0.12}_{-0.083}$ & $2.42 \pm 0.59$ \\
       Centaurus & $1.209 \pm 0.014$ & $13.04 \pm 0.42$ & $0.313 \pm 0.017$ & $0.294 \pm 0.029$ & $0.255 \pm 0.032$ & $<0.059$ & $0.233 \pm 0.016$ & $1.75 \pm 0.11$ \\
          HCG 62 & $0.7472 \pm 0.0079$ & $3^{+3}_{-2}$ & $0.305^{+0.086}_{-0.071}$ & $0.81^{+0.33}_{-0.26}$ & $0.61^{+0.27}_{-0.21}$ & $9^{+7}_{-5}$ & $0.61^{+0.22}_{-0.17}$ & $1.88^{+0.74}_{-0.59}$ \\
      Abell 1650 & $3.44 \pm 0.58$ & $<0.89$ & $0.37^{+0.20}_{-0.14}$ & $0.51^{+0.26}_{-0.20}$ & $<0.30$ & $3^{+2}_{-1}$ & $0.236^{+0.11}_{-0.089}$ & $<0.51$ \\
        NGC 5044 & $0.7589 \pm 0.0097$ & $<1$ & $0.317^{+0.070}_{-0.058}$ & $0.56 \pm 0.16$ & $0.81 \pm 0.16$ & $<5$ & $0.559 \pm 0.094$ & $2.32 \pm 0.51$ \\
      Abell 3558 & $2.34^{+0.87}_{-0.33}$ & $6 \pm 1$ & $0.41^{+0.25}_{-0.16}$ & $<0.26$ & $<0.53$ & $<0.39$ & $0.203^{+0.20}_{-0.075}$ & $<0.91$ \\
 RX J1347.5-1145 & $4.32^{+3}_{-0.50}$ & $4 \pm 2$ & $0.21^{+0.32}_{-0.15}$ & $0.68^{+1}_{-0.26}$ & $0.58^{+1}_{-0.35}$ & $<0.68$ & $0.193^{+1}_{-0.085}$ & $<7$ \\
      Abell 1795 & $2.76^{+0.33}_{-0.11}$ & $1.46 \pm 0.61$ & $0.307^{+0.069}_{-0.052}$ & $0.553^{+0.16}_{-0.097}$ & $0.32 \pm 0.15$ & $<0.85$ & $0.319^{+0.079}_{-0.039}$ & $0.66 \pm 0.47$ \\
      Abell 3581 & $1.3120 \pm 0.0090$ & $6.42 \pm 0.50$ & $0.262 \pm 0.020$ & $0.269 \pm 0.047$ & $0.319 \pm 0.047$ & $<0.17$ & $0.272 \pm 0.014$ & $0.74 \pm 0.14$ \\
      Abell 1991 & $1.629 \pm 0.035$ & $4 \pm 1$ & $0.357 \pm 0.067$ & $0.33 \pm 0.11$ & $0.20 \pm 0.13$ & $<0.31$ & $0.393 \pm 0.044$ & $<0.66$ \\
     E 1455+2232 & $2.73^{+2}_{-0.36}$ & $<3$ & $0.35^{+0.64}_{-0.15}$ & $<0.71$ & $<0.67$ & $<0.54$ & $0.217^{+0.72}_{-0.078}$ & $<5$ \\
        NGC 5813 & $0.6259 \pm 0.0073$ & $<0.41$ & $0.267 \pm 0.029$ & $0.301 \pm 0.082$ & $0.545 \pm 0.094$ & $<0.59$ & $0.516^{+0.091}_{-0.045}$ & $1.37 \pm 0.31$ \\
RXC J1504.1-0248 & $3.19^{+1}_{-0.45}$ & $9 \pm 1$ & $0.300^{+0.15}_{-0.098}$ & $0.30^{+0.35}_{-0.12}$ & $0.42^{+0.35}_{-0.16}$ & $0.263^{+0.19}_{-0.086}$ & $0.175^{+0.22}_{-0.056}$ & $1.03^{+1}_{-0.56}$ \\
        NGC 5846 & $0.6260 \pm 0.0089$ & $<0.91$ & $0.64^{+0.71}_{-0.13}$ & $0.80^{+0.90}_{-0.18}$ & $0.51 \pm 0.23$ & $<4$ & $0.90^{+0.96}_{-0.17}$ & $4.67^{+3}_{-0.73}$ \\
      Abell 2029 & $3.431^{+0.20}_{-0.085}$ & $3.23 \pm 0.26$ & $0.272 \pm 0.032$ & $0.321 \pm 0.066$ & $0.341 \pm 0.079$ & $0.85^{+0.60}_{-0.48}$ & $0.231^{+0.030}_{-0.020}$ & $0.94 \pm 0.27$ \\
      Abell 2052 & $1.678 \pm 0.015$ & $6.45 \pm 0.40$ & $0.326 \pm 0.022$ & $0.355 \pm 0.046$ & $0.261 \pm 0.052$ & $<0.083$ & $0.306 \pm 0.014$ & $<0.19$ \\
          MKW 3s & $2.728^{+0.23}_{-0.099}$ & $1.51 \pm 0.67$ & $0.440 \pm 0.080$ & $0.52 \pm 0.15$ & $<0.42$ & $<0.53$ & $0.505^{+0.086}_{-0.060}$ & $<1$ \\
      Abell 2063 & $3.13^{+0.87}_{-0.69}$ & $6 \pm 2$ & $0.52^{+0.38}_{-0.27}$ & $<1$ & $<1$ & $<3$ & $0.61^{+0.48}_{-0.29}$ & $<4$ \\
      Abell 2199 & $2.00 \pm 0.11$ & $<1$ & $0.316 \pm 0.052$ & $0.38 \pm 0.13$ & $0.57 \pm 0.16$ & $<3$ & $0.315 \pm 0.059$ & $0.95 \pm 0.44$ \\
      Abell 2204 & $3.14 \pm 0.18$ & $8.05 \pm 0.41$ & $0.286 \pm 0.045$ & $0.481 \pm 0.076$ & $0.373 \pm 0.081$ & $0.365 \pm 0.069$ & $0.286 \pm 0.038$ & $<0.19$ \\
      Hercules A & $2.43 \pm 0.33$ & $7^{+1}_{-1}$ & $0.51^{+0.16}_{-0.13}$ & $0.28^{+0.16}_{-0.13}$ & $<0.31$ & $0.46 \pm 0.16$ & $0.237 \pm 0.078$ & $0.85 \pm 0.62$ \\
 RX J1720.1+2638 & $3.50 \pm 0.63$ & $6^{+1}_{-1}$ & $0.31 \pm 0.11$ & $0.32 \pm 0.17$ & $0.37 \pm 0.21$ & $0.53^{+0.24}_{-0.20}$ & $0.35^{+0.18}_{-0.12}$ & $2^{+1}_{-1}$ \\
RXC J2014.8-2430 & $3.75^{+3}_{-0.57}$ & $12 \pm 2$ & $0.32^{+0.37}_{-0.14}$ & $0.61^{+1}_{-0.23}$ & $<0.27$ & $0.24^{+0.48}_{-0.20}$ & $0.45^{+1}_{-0.15}$ & $<0.75$ \\
 RX J2129.6+0005 & $3.35 \pm 0.67$ & $2^{+1}_{-1}$ & $0.241^{+0.13}_{-0.099}$ & $<0.21$ & $0.64^{+0.37}_{-0.30}$ & $0.37^{+0.28}_{-0.23}$ & $0.26^{+0.15}_{-0.10}$ & $<1$ \\
RXC J2149.1-3041 & $2.11^{+0.26}_{-0.19}$ & $1 \pm 1$ & $0.37 \pm 0.12$ & $0.57 \pm 0.22$ & $0.59 \pm 0.25$ & $1.07^{+0.68}_{-0.53}$ & $0.307^{+0.11}_{-0.079}$ & $<0.21$ \\
     Abell S1101 & $2.179^{+0.13}_{-0.049}$ & $0.82 \pm 0.40$ & $0.258^{+0.032}_{-0.027}$ & $0.335 \pm 0.069$ & $0.271 \pm 0.076$ & $0.41 \pm 0.24$ & $0.379^{+0.044}_{-0.028}$ & $0.55 \pm 0.21$ \\
      Abell 2597 & $2.62 \pm 0.11$ & $1.44 \pm 0.39$ & $0.353 \pm 0.042$ & $0.52 \pm 0.10$ & $0.324 \pm 0.096$ & $<0.13$ & $0.366 \pm 0.043$ & $0.47 \pm 0.28$ \\
      Abell 2626 & $2.66 \pm 0.25$ & $2 \pm 1$ & $0.244^{+0.12}_{-0.096}$ & $0.78 \pm 0.29$ & $0.57 \pm 0.32$ & $1^{+1}_{-1}$ & $0.64 \pm 0.16$ & $2 \pm 1$ \\
      Klemola 44 & $2.69 \pm 0.43$ & $1 \pm 1$ & $0.47 \pm 0.14$ & $0.30 \pm 0.24$ & $<0.17$ & $<5$ & $0.38 \pm 0.14$ & $<0.93$ \\
      Abell 2667 & $3 \pm 1$ & $<2$ & $0.20^{+0.23}_{-0.16}$ & $0.33^{+0.31}_{-0.22}$ & $<0.69$ & $0.61^{+0.44}_{-0.29}$ & $0.32^{+0.37}_{-0.18}$ & $2^{+2}_{-1}$ \\
      Abell 4059 & $2.21 \pm 0.15$ & $2 \pm 1$ & $0.44 \pm 0.10$ & $0.45 \pm 0.24$ & $0.29 \pm 0.26$ & $<0.90$ & $0.51^{+0.14}_{-0.11}$ & $<1$ \\

  \hline
\end{longtable}

\begin{longtable}{lcccccc}
  \caption{Upper limits on broadening and redshifts. Shown are
    measured redshifts for each object (and $1\sigma$ statistical
    errors), the upper limit on the broadening of the emission lines
    for each object (at the 90 per cent level), the upper limits for
    cooler component of two temperature model (marked 2T) and the
    upper limits using response matrices
    with 10 per cent narrower and broader LSF components.}\\
  \label{tab:vel}
  Cluster & Redshift & Limit & MCMC Limit & 2T Limit & Narrow LSF
  Limit & Broad LSF Limit \\ & & (\kmps) & (\kmps) & (\kmps) & (\kmps) & (\kmps) \\ \hline
  Abell 133 & $0.05720 \pm 0.00051$ & $<1200$ & $<1200$ & & $<1200$ & $<1200$ \\
NGC 499 & $0.01636 \pm 0.00091$ & $<2800$ & $<2800$ & & $<2800$ & $<2800$ \\
NGC 533 & $0.01869 \pm 0.00025$ & $<600$ & $<570$ & & $<630$ & $<580$ \\
Abell 262 & $0.01742 \pm 0.00027$ & $<960$ & $<1000$ & $<640$ & $<980$ & $<940$ \\
NGC 777 & $0.01738 \pm 0.00036$ & $<1200$ & $<1200$ & & $<1300$ & $<1200$ \\
Abell 383 & $0.19070 \pm 0.00089$ & $<1100$ & $<1200$ & & $<1200$ & $<1100$ \\
NGC 1316 & $0.00617 \pm 0.00014$ & $<650$ & $<660$ & & $<670$ & $<620$ \\
NGC 1399 & $0.00437 \pm 0.00012$ & $<570$ & $<580$ & & $<590$ & $<540$ \\
2A 0335+096 & $0.03634 \pm 0.00016$ & $<900$ & $<930$ & $<1500$ & $<920$ & $<870$ \\
NGC 1404 & $0.00723 \pm 0.00021$ & $<1100$ & $<1100$ & & $<1100$ & $<1100$ \\
NGC 1550 & $0.0132 \pm 0.0010$ & $<2500$ & $<2900$ & & $<2500$ & $<2500$ \\
RBS 540 & $0.03848 \pm 0.00031$ & $<700$ & $<780$ & & $<720$ & $<680$ \\
Abell 496 & $0.03295 \pm 0.00020$ & $<1100$ & $<1100$ & & $<1100$ & $<1000$ \\
RXC J0605.8-3518 & $0.14110^{+0.00078}_{-0.00063}$ & $<690$ & $<700$ & & $<700$ & $<670$ \\
NGC 2300 & $0.00769 \pm 0.00033$ & $<890$ & $<970$ & & $<910$ & $<870$ \\
MS 0735.6+7421 & $0.21730^{+0.00059}_{-0.00047}$ & $<670$ & $<830$ & & $<680$ & $<660$ \\
PKS 0745-19 & $0.10140 \pm 0.00061$ & $<1400$ & $<1900$ & & $<1400$ & $<1400$ \\
Hydra A & $0.05405 \pm 0.00030$ & $<820$ & $<840$ & & $<830$ & $<800$ \\
RBS 797 & $0.35410 \pm 0.00062$ & $<450$ & $<950$ & & $<440$ & $<460$ \\
Zw3146 & $0.28870 \pm 0.00027$ & $<480$ & $<500$ & & $<500$ & $<460$ \\
Abell 1068 & $0.13890 \pm 0.00045$ & $<620$ & $<680$ & & $<660$ & $<590$ \\
RXC J1044.5-0704 & $0.13200 \pm 0.00053$ & $<620$ & $<730$ & & $<640$ & $<600$ \\
NGC 3411 & $0.01490 \pm 0.00037$ & $<1300$ & $<1400$ & & $<1300$ & $<1300$ \\
RXC J1141.4-1216 & $0.12110 \pm 0.00053$ & $<810$ & $<1100$ & & $<830$ & $<780$ \\
Abell 1413 & $0.14360 \pm 0.00098$ & $<1400$ & $<2200$ & & $<1400$ & $<1400$ \\
MKW 4 & $0.01975 \pm 0.00097$ & $<1900$ & $<1900$ & & $<1900$ & $<1900$ \\
NGC 4261 & $0.00755 \pm 0.00013$ & $<320$ & $<330$ & & $<360$ & $<290$ \\
NGC 4325 & $0.02534 \pm 0.00029$ & $<950$ & $<960$ & & $<970$ & $<930$ \\
NGC 4472 & $0.00294 \pm 0.00012$ & $<700$ & $<700$ & & $<720$ & $<680$ \\
NGC 4552 & $0.00180 \pm 0.00031$ & $<1100$ & $<1100$ & & $<1100$ & $<1100$ \\
NGC 4636 & $0.00284 \pm 0.00013$ & $<1000$ & $<1000$ & $<380$ & $<1000$ & $<1000$ \\
NGC 4649 & $0.00379 \pm 0.00014$ & $<750$ & $<750$ & & $<760$ & $<720$ \\
Centaurus & $0.01041 \pm 0.00015$ & $<1100$ & $<1100$ & $<540$ & $<1100$ & $<1000$ \\
HCG 62 & $0.01472 \pm 0.00013$ & $<770$ & $<780$ & & $<780$ & $<750$ \\
Abell 1650 & $0.08445^{+0.00067}_{-0.00035}$ & $<640$ & $<900$ & & $<650$ & $<630$ \\
NGC 5044 & $0.00900 \pm 0.00033$ & $<1500$ & $<1500$ & & $<1500$ & $<1500$ \\
Abell 3558 & $0.0513 \pm 0.0043$ & $<4000$ & $<6700$ & & $<4000$ & $<4000$ \\
RX J1347.5-1145 & $0.4493^{+0.0022}_{-0.0012}$ & $<1800$ & $<3200$ & & $<1800$ & $<1800$ \\
Abell 1795 & $0.06447 \pm 0.00040$ & $<1000$ & $<1100$ & & $<1100$ & $<990$ \\
Abell 3581 & $0.02147 \pm 0.00026$ & $<1300$ & $<1300$ & $<730$ & $<1300$ & $<1300$ \\
Abell 1991 & $0.05813 \pm 0.00027$ & $<460$ & $<450$ & $<920$ & $<480$ & $<430$ \\
E 1455+2232 & $0.25750 \pm 0.00084$ & $<830$ & $<980$ & & $<690$ & $<820$ \\
NGC 5813 & $0.00827 \pm 0.00027$ & $<1400$ & $<1500$ & & $<1400$ & $<1400$ \\
RXC J1504.1-0248 & $0.21720^{+0.00083}_{-0.00062}$ & $<1400$ & $<1900$ & & $<1400$ & $<1400$ \\
NGC 5846 & $0.00626 \pm 0.00020$ & $<870$ & $<860$ & & $<880$ & $<850$ \\
Abell 2029 & $0.07872 \pm 0.00026$ & $<600$ & $<610$ & & $<620$ & $<580$ \\
Abell 2052 & $0.03443 \pm 0.00020$ & $<1100$ & $<1100$ & $<820$ & $<1100$ & $<1100$ \\
MKW 3s & $0.04352 \pm 0.00094$ & $<2500$ & $<2600$ & & $<2500$ & $<2500$ \\
Abell 2063 & $0.0331 \pm 0.0020$ & $<2000$ & $<3400$ & & $<2000$ & $<2000$ \\
Abell 2199 & $0.03094 \pm 0.00036$ & $<820$ & $<880$ & & $<840$ & $<790$ \\
Abell 2204 & $0.15240 \pm 0.00020$ & $<460$ & $<500$ & & $<490$ & $<430$ \\
Hercules A & $0.15500^{+0.00088}_{-0.00070}$ & $<1300$ & $<1400$ & & $<1300$ & $<1200$ \\
RX J1720.1+2638 & $0.15990 \pm 0.00054$ & $<820$ & $<890$ & & $<860$ & $<780$ \\
RXC J2014.8-2430 & $0.15480 \pm 0.00086$ & $<920$ & $<1400$ & & $<910$ & $<920$ \\
RX J2129.6+0005 & $0.23410 \pm 0.00051$ & $<610$ & $<3100$ & & $<520$ & $<880$ \\
RXC J2149.1-3041 & $0.12200^{+0.00046}_{-0.00030}$ & $<320$ & $<400$ & & $<320$ & $<310$ \\
Abell S1101 & $0.05563 \pm 0.00026$ & $<1000$ & $<1100$ & $<510$ & $<1100$ & $<1000$ \\
Abell 2597 & $0.08252 \pm 0.00023$ & $<660$ & $<670$ & & $<680$ & $<640$ \\
Abell 2626 & $0.05415 \pm 0.00057$ & $<880$ & $<920$ & & $<900$ & $<860$ \\
Klemola 44 & $0.02819 \pm 0.00055$ & $<970$ & $<3800$ & & $<990$ & $<950$ \\
Abell 2667 & $0.23240^{+0.00067}_{-0.00055}$ & $<410$ & $<810$ & & $<400$ & $<420$ \\
Abell 4059 & $0.04873 \pm 0.00056$ & $<1200$ & $<1400$ & & $<1200$ & $<1200$ \\

  \hline
\end{longtable}

\twocolumn
\begin{table}
  \caption{\emph{Chandra} datasets examined for spectral
    mapping. Exposure shows the cleaned total exposure time. S/N shows
    the minimum signal to noise ratio used to create the bins in the
    binning (square to get the approximate number of counts per bin).}
  \label{tab:chandraobs}
  \begin{tabular}{lp{2.6cm}cc}
    \hline
    Object     & Observations & Exposure (ks) & S/N \\ \hline
    Abell 133  & 2203 & 34 & 30 \\
    Abell 383  & 2320, 2321 & 38 & 30 \\
    Abell 496  & 931, 3361, 4976 & 88 & 40 \\
    Abell 1068 & 1652 & 26 & 30 \\
    Abell 1650 & 5822, 5823, 6356, 6358, 7242 & 164 & 50\\
    Abell 1795 & 493, 494, 3666, 5286, 5287, 5288, 5289, 5290, 6159,
    6160, 6161, 6162, 6163 & 91 & 32 \\
    Abell 1991 & 3193 & 37 & 20\\
    Abell 2063 & 5795, 6262, 6263 & 41 & 30 \\
    Abell 2204 & 499, 6104, 7940 & 97 & 45 \\
    Abell 2597 & 922 6934 7329 & 141 & 30 \\
    Abell 2626 & 3192 & 24 & 20 \\
    Abell 2667 & 2214 & 10 & 20 \\
    Abell 3581 & 1650 & 7 & 15 \\
    Abell 4059 & 897, 5785 & 129 & 30 \\
    E 1455.0+2232 & 543, 4192, 7709 & 108 & 20\\
    HCG 62     & 921 & 48 & 22 \\
    Hercules A & 1625, 5796, 6257 & 112 & 30\\
    Hydra A    & 4969, 4970 & 182 & 100 \\
    Klem 44    & 4188, 4992 & 39 & 30\\
    MKW 4      & 3234 & 29  & 30 \\
    MS 0735.6+7421 & 4197 & 45 & 20 \\
    NGC 533    & 2880 & 36 & 20 \\
    NGC 1316   & 2022 & 28 & 20 \\
    NGC 1399   & 319, 9530 & 115 & 40 \\
    NGC 4261   & 834, 9569 & 134 & 30 \\
    NGC 5044   & 798, 9399 & 103 & 30 \\
    NGC 5813   & 5907, 9517 & 147 & 30 \\
    PKS 0745-19& 508, 2427, 6103, 7694 & 44 & 32 \\
    RBS 540    & 4183 & 10 & 20 \\
    RBS 797    & 7902 & 38 & 20 \\
    RX J1347.5-1145 & 3592 & 57 & 30 \\
    RX J1720.1+2638 & 3224, 4361 & 44 & 30 \\
    RX J2129.6+0005 & 552, 9370 & 40 & 30 \\
    Zw 3146    & 909, 9371 & 84 & 30 \\
    \hline
  \end{tabular}
\end{table}

\begin{table}
  \caption{Measured line widths and line widths from spectra simulated
    using \emph{Chandra} maps in the absence of turbulence. 
    The difference column shows the difference between best fitting
    velocities of the observed and simulated spectra. The limit column
    shows the 90 per cent upper limit on the measured velocity after
    subtracting the predicted velocity.
    Uncertainties are $1\sigma$ in this table.
    Objects indicated with $^*$ show the results using an MCMC analysis.}
  \label{tab:linewidths}
  \begin{tabular}{lcccc}
    \hline
    Object        & Real                 & Predicted    & Difference          & Limit\\
                  & (\kmps)              & (\kmps)      & (\kmps)             & (\kmps)\\     
    \hline
    Abell 133     & $910 \pm 170$        & $771 \pm 17$ & $140 \pm 170$       & 260 \\
    Abell 383     & $670 \pm 250$        & $377 \pm 26$ & $290 \pm 250$       & 700 \\
    Abell 496     & $1050 \pm 70$        & $1065 \pm 6$ & $-15 \pm 70$        & 100 \\
    Abell 1068    & $50^{+405}_{-50}$    & $536 \pm 23$ & $-490^{+405}_{-55}$ & 180 \\
    Abell 1650$^*$& $45^{+790}_{-45}$    & $1034 \pm 70$& $-990^{+790}_{-45}$ & 310\\
    Abell 1795    & $1040 \pm 195$       & $1030 \pm 20$& $10 \pm 195$        & 330 \\
    Abell 2063    & $1530_{-405}^{+460}$ & $1780 \pm 65$& $-250^{+460}$       & 510 \\
    Abell 2204    & $250^{+110}_{-125}$  & $282 \pm 13$ & $-30^{+110}_{-125}$ & 150 \\
    Abell 2597    & $510 \pm 75$         & $674 \pm 12$ & $-160 \pm 75$       & -40 \\
    Abell 2626    & $705^{+210}_{-190}$  & $955 \pm 25$ & $-250^{+210}_{-190}$& 100 \\
    Abell 2667    & $0^{+240}$           & $374 \pm 40$ & $-370^{+240}$       & 25  \\
    Abell 4059    & $1050^{+205}_{-175}$ & $1350 \pm 20$& $-300^{+205}$       & 40  \\
    E 1455.0+2232 & $0^{+460}$           & $375 \pm 33$ & $-375^{+460}$       & 380 \\
    HCG 62        & $710 \pm 50$         & $749 \pm 4$  & $-40 \pm 50$        & 40  \\
    Hercules A    & $770^{+300}_{-230}$  & $573 \pm 20$ & $200^{+300}_{-230}$ & 700 \\
    Hydra A       & $750 \pm 115$        & $867 \pm 12$ & $-120 \pm 115$      & 70  \\
    Klem 44$^*$   & $3660_{-720}^{+960}$ & $1892 \pm 50$& $2350_{-720}^{+960}$& 1160\\
    MKW 4         & $1320_{-415}^{+345}$ & $970 \pm 20$ & $350_{-415}^{+345}$ & 920 \\
    MS 0735.6+7421& $300 \pm 300$        & $970 \pm 40$ & $-470 \pm 300$      & 25 \\
    NGC 533       & $530 \pm 105$        & $541 \pm 8$  & $-10 \pm 105$       & 160 \\
    NGC 1316      & $650 \pm 60$         & $685 \pm 6$  & $-145 \pm 65$       & 65 \\
    NGC 1399      & $590 \pm 60$         & $545 \pm 4$  & $45 \pm 60$         & 140 \\
    NGC 5044      & $1530^{+120}_{-95}$  & $1760 \pm 16$& $-230^{+120}_{-95}$ & -32 \\
    NGC 5813      & $1605 \pm 95$        & $1540 \pm 6$ & $65 \pm 95$         & 220 \\
    PKS 0745-19$^*$& $910^{+600}_{-880}$  & $790 \pm 65$ & $120^{+600}_{-880}$  & 1110 \\
    RBS 540       & $545 \pm 165$        & $700 \pm 15$ & $-155 \pm 165$      & 120 \\
    RBS 797$^*$     & $0^{+535}$           & $673^{+30}_{-55}$ & $-675^{+535}$  & 210\\
    RX J1347.5-1145$^*$&$1320^{+770}_{-430}$&$200 \pm 100$& $1120^{+770}_{-430}$& 410 \\
    RX J1720.1+2638 & $530^{+290}_{-250}$& $431 \pm 25$ & $100^{+290}_{-250}$ & 580 \\
    RX J2129.6+0005$^*$& $0^{+1530}$        & $470 \pm 45$ &  $-470^{+1530}$      & 2150 \\
    Zw 3146       & $310 \pm 100$        & $320 \pm 10$ & $-10 \pm 100$       & 155 \\
    \hline
  \end{tabular}
\end{table}
\normalsize

\end{document}